\documentstyle{article}  
\newlength{\extralineskip}
\newcommand{\beq}{\begin{equation}}
\newcommand{\eeq}{\end{equation}}
\newcommand{\bd}{\begin{displaymath}}
\newcommand{\ed}{\end{displaymath}}

\def\bea{\begin{eqnarray}}
\def\eea{\end{eqnarray}}

\def\ba{\beq\new\begin{array}{c}}
\def\ea{\end{array}\eeq}

\def\inbar{\,\vrule height1.5ex width.4pt depth0pt}
\def\IC{\relax\hbox{$\inbar\kern-.3em{\rm C}$}}
\def\IR{\relax{\rm I\kern-.18em R}}

\def\IN{\relax{\rm I\kern-.15em N}}

\makeatletter
\newdimen\normalarrayskip              
\newdimen\minarrayskip                 
\normalarrayskip\baselineskip
\minarrayskip\jot
\newif\ifold             \oldtrue            \def\new{\oldfalse}
\def\arraymode{\ifold\relax\else\displaystyle\fi} 
\def\@arrayskip{\ifold\baselineskip\z@\lineskip\z@
     \else
     \baselineskip\minarrayskip\lineskip2\minarrayskip\fi}
\def\@arrayclassz{\ifcase \@lastchclass \@acolampacol \or
\@ampacol \or \or \or \@addamp \or
   \@acolampacol \or \@firstampfalse \@acol \fi
\edef\@preamble{\@preamble
  \ifcase \@chnum
     \hfil$\relax\arraymode\@sharp$\hfil
     \or $\relax\arraymode\@sharp$\hfil
     \or \hfil$\relax\arraymode\@sharp$\fi}}
\def\@array[#1]#2{\setbox\@arstrutbox=\hbox{\vrule
     height\arraystretch \ht\strutbox
     depth\arraystretch \dp\strutbox
     width\z@}\@mkpream{#2}\edef\@preamble{\halign \noexpand\@halignto
\bgroup \tabskip\z@ \@arstrut \@preamble \tabskip\z@ \cr}%
\let\@startpbox\@@startpbox \let\@endpbox\@@endpbox
  \if #1t\vtop \else \if#1b\vbox \else \vcenter \fi\fi
  \bgroup \let\par\relax
  \let\@sharp##\let\protect\relax
  \@arrayskip\@preamble}

\begin{document}
\thispagestyle{empty}

\begin{center}
{\huge \bf On the Correspondence Between the 
Strongly Coupled 2-Flavor Lattice 
Schwinger Model and the
Heisenberg Antiferromagnetic Chain}\\
\vskip 0.3 truein
{\bf F. Berruto$^{(a)}$, G. Grignani$^{(a)}$, G. W. Semenoff$^{(b)}$ 
and P. Sodano$^{(a)}$}
\vskip 0.3truein
$(a)$ {\it Dipartimento di Fisica and Sezione
I.N.F.N., Universit\`a di Perugia, Via A. Pascoli I-06123 Perugia,
Italy}\\
\vskip 1.truecm
$(b)$ {\it Department of Physics and Astronomy\\University of British 
Columbia\\
6224 Agricultural Road\\Vancouver, British Columbia, Canada V6T 1Z1}\\
\vskip 1.0truein
DFUPG-190-98, UBC/GS-6-98
\vskip 1.0truein
{\bf Abstract}
\vskip 0.3truein
\end{center}
We study the strong coupling limit of the 2-flavor massless Schwinger
model on a lattice using staggered fermions and the Hamiltonian
approach to lattice gauge theories.  Using the correspondence between
the low-lying states of the 2-flavor strongly coupled lattice
Schwinger model and the antiferromagnetic Heisenberg chain established
in a previous paper, we explicitly compute the mass spectrum of this
lattice gauge model: we identify the low-lying excitations of the
Schwinger model with those of the Heisenberg model and compute the
mass gaps of other excitations in terms of vacuum expectation values
(v.e.v.'s) of powers of the Heisenberg Hamiltonian and spin-spin
correlation functions.  We find a satisfactory agreement with the
results of the continuum theory already at the second order in the
strong coupling expansion.  We show that the pattern of symmetry
breaking of the continuum theory is well reproduced by the lattice
theory; we see indeed that in the lattice theory the isoscalar
$\left<\bar\psi\psi\right>$ and isovector $\left<
\bar\psi \sigma^a\psi\right>$ chiral condensates are zero to every
order in the strong coupling expansion.  In addition, we find that the
chiral condensate $<\overline{\psi}_{L}^{(2)}\overline{\psi}_{L}^{(1)}
\psi_{R}^{(1)}\psi_{R}^{(2)}>$ is non zero also on the lattice; 
this is the relic in this lattice model of the axial anomaly in the
continuum theory.

We compute the v.e.v.'s of the spin-spin correlators of the Heisenberg
model which are pertinent to the calculation of the mass spectrum and
obtain an explicit construction of the lowest lying states for finite
size Heisenberg antiferromagnetic chains.
\newpage
\setcounter{page}1

\section{Introduction}

One of the analytical approaches to the study of gauge theories with
confining spectra is the strong coupling expansion. In the strong
coupling limit, confinement is explicit, the confining string is a
stable object\cite{wilson} and some other qualitative features of the
spectrum are easily obtained. Formulation of the strong coupling
expansion requires a gauge invariant ultraviolet cutoff, which is most
conveniently implemented using a lattice regularization.

One of the
most difficult problems of the strong coupling approach to lattice
gauge theory is its extrapolation to the continuum limit, which usually
occurs at weak coupling.  One symptom of this problem is that many choices 
of strong coupling theory produce identical continuum physics. 
In spite of this difficulty, there are 
strong coupling computations which claim 
some degree of success~\cite{jones,kogut,noi,stein,noi2}. It is well known that between strongly coupled lattice
gauge theories and quantum spin systems there is an intimate
relationship
\cite{kogu,smit, semen} which is most useful 
to analyze chiral symmetry breaking in a variety of lattice models.  

A useful test of strong coupling expansions in lattice gauge theories
is the study of models whose solution in the continuum is known even
in the strong coupling regime.  For example, the 1-flavor Schwinger
model has been studied using the Hamiltonian lattice field theory
within the strong coupling expansion~\cite{kogut,noi}; several
relevant parameters, analytically computed on the lattice, are shown
to be in good agreement with those of the exact continuum
solution~\cite{noi}. Furthermore, issues involving the realization of
chiral symmetry on the lattice can be studied
systematically~\cite{noi}. The 1-flavor Schwinger model has also been used as an example of 
the fact that quantum link models may reproduce the physics of conventional Hamiltonian 
lattice gauge theories~\cite{chandra}. The physically more interesting case of the
2-flavor Schwinger model has been studied in~\cite{stein, noi2};
in addition to the issues of spectrum and chiral symmetry breaking,
one sees that the strong coupling limit of this model is
mapped~\cite{hoso, noi2} on a relevant quantum spin model $-$ the
one-dimensional spin-$1/2$ quantum Heisenberg
antiferromagnet. This correspondence is very useful since the ground
state of the antiferromagnetic chain is known~\cite{bethe} and its
energy has been computed in~\cite{hulten}; moreover, the complete
spectrum has been determined by Faddeev and Takhtadzhyan
\cite{faddeev} using the algebraic Bethe ansatz.

There are by now many hints at a correspondence between quantized
gauge theories and quantum spin models, aimed at analyzing new phases
relevant for condensed matter systems \cite{anderson}. Recently,
Laughlin has argued that there is an analogy between the spectral data
of gauge theories and strongly correlated electron systems
\cite{BL}. Moreover, certain spin ladders have been related to the
2-flavor Schwinger model in Ref.~\cite{hoso} where a relation between
the physical parameters of the spin and the gauge systems was also
found.

In this context, the mapping between the strongly coupled 2-flavor
Schwinger model and the quantum Heisenberg antiferromagnetic chain
provides a concrete computational scheme in which the issue of the
correspondence between quantized gauge theories and quantum spin
models may be investigated. Because of the dimensionality of the
coupling constant in (1+1)-dimensions, the infrared behavior is
governed by the strong coupling limit, and it is tempting to
conjecture the existence of an exact correspondence between the
infrared limits of the Heisenberg and 2-flavor Schwinger models.

In this paper we revisit the strong coupling limit of the 2-flavor lattice 
Schwinger model in  the Hamiltonian formalism with staggered fermions.
Using the results of Ref.\cite{noi2} we analyze in detail the 
spectrum of the model: the gauge theory vacuum is  
the ground state of the Heisenberg antiferromagnetic chain and 
we find that the low-lying excitations of the 
gauge model have the same quantum numbers of the states of 
the spin chain. 
In addition we compute the masses 
of the excitations to the second order in the strong coupling expansion; 
as pointed out in \cite{noi2}, the 
computation needs the knowledge of some spin-spin correlators of the quantum 
Heisenberg antiferromagnetic chain, which we explicitly 
compute~\cite{taka,korepin}. Our analysis hints to the existence of 
two other massive isotriplet states in addition to the expected pseudoscalar
massive isosinglet.

We analyze then the pattern of symmetry breaking in the lattice theory.
Even though the continuum axial symmetry is broken explicitly by the
staggered fermions, a discrete axial symmetry remains, it corresponds 
to a chiral rotation of $\pi/2$ and appears in the lattice theory as a 
translation by one lattice site. Since the ground state of the 
Heisenberg chain is unique and translationally invariant - 
at variance with the one flavor Schwinger model - in the 2-flavor 
model the discrete axial symmetry is unbroken.
The chiral anomaly in this model is realized on the lattice via the 
explicit breaking of the $U_{A}(1)$-symmetry induced by the staggered 
fermions.
Therefore, the pattern of symmetry breaking on the lattice
reproduces faithfully the one of the continuum theory.

We also compute the chiral condensates in the strongly coupled 
2-flavor lattice Schwinger model showing that the results
of the lattice are in agreement with those of the continuum.
In the continuum there is neither an isoscalar 
$\left<\bar\psi\psi\right>$ nor an isovector 
$\left< \bar\psi \sigma^a\psi\right>$ chiral condensate, 
since this is forbidden by the Coleman 
theorem~\cite{colemant}. We show that on the lattice these 
fermion condensates are also zero to all the 
orders in the strong coupling expansion. 
We find that $-$ just as in the continuum theory $-$ 
the non-vanishing chiral condensate is given by the 
v.e.v. $<\overline{\psi}_{L}^{(2)}\overline{\psi}_{L}^{(1)}
\psi_{R}^{(1)}\psi_{R}^{(2)}>$, 
which we compute up to the 
second order in the strong coupling expansion. 

We finally use both the mass spectrum and the 
non-vanishing chiral condensate to 
compute, by means of suitable ``Pad\'e approximants", 
the physical parameters of this lattice model 
and then compare their numerical values with the exact results
of the continuum theory.   

In section 2 we review known results \cite{coleman,hara} of the 
continuum 2-flavor Schwinger model and define its Hamiltonian 
lattice version. We also identify the lattice counterparts of the relevant 
symmetries used in \cite{coleman} for the description of the spectrum 
of the continuum theory. 

In section 3 we expand the results of Ref.\cite{noi2} and set up the 
formalism needed for 
the strong coupling analysis of the lattice 2-flavor 
Schwinger model. We show that the ground and excited states of 
the antiferromagnetic 
Heisenberg model have the spectrum and quantum numbers which are 
expected for the ground state and 
massless excitations of the 2-flavor Schwinger model. 

In section 4 we construct the operators that create the massive 
excitations when acting on the ground state. 
We set up the strong coupling expansion and compute the 
corrections to the energies of the ground state and of the low-lying massive 
bosonic excitations. 
Subtracting the energy of the ground state from those of the 
excitations defines the 
lattice meson masses. 

In section 5 we show that both the isoscalar and the isovector 
chiral condensates are zero to every perturbative order in the 
strong coupling expansion due to the translational invariance of the 
strong coupling ground state. We also compute, up to the 
second order in the strong coupling expansion, the chiral condensate 
$<\overline{\psi}_{L}^{(2)}\overline{\psi}_{L}^{(1)}\psi_{R}^{(1)}
\psi_{R}^{(2)}>$, 
which is the 
order parameter for the breaking of the $U_{A}(1)$ symmetry. 

Section 6 is devoted to the comparison of 
the lattice results with the continuum theory; there we shall see 
that the lattice theory, properly extrapolated to the 
continuum via the use of Pad\'e approximants, well reproduces the 
parameters of the continuum theory already at the second order of the 
strong coupling expansion.

Section 7 is devoted to some concluding remarks.

In the appendix A, after reviewing known results about the Bethe 
Ansatz solution \cite{faddeev} of the antiferromagnetic Heisenberg 
chain, we study explicitly the complete spectrum of finite size 
Heisenberg antiferromagnetic chains of 4 and 6 sites in order
to compare it to the Bethe Ansatz solution obtained in the thermodynamic 
limit. 
We write down explicitly the ground states of 4,6 and 8 site chains 
and we find the that, even for these very small systems, the results 
exhibited in \cite{faddeev} are very well reproduced. 
This provides also an useful intuitive picture of the strong coupling 
ground state of the 2-flavor lattice Schwinger model.

In appendix B we comment on the computation of the spin-spin correlator
of the Heisenberg chain and provide a link between the results given 
in~\cite{taka} and the method used in~\cite{korepin}. 

\section{Two flavor Schwinger model in the continuum and on the lattice}

The action of the $1+1$-dimensional electrodynamics with two charged 
Dirac spinor fields is
\begin{equation}
S = \int d^{2} x\left[\sum_{a=1}^{2} \overline{\psi}_{a}
(i\gamma_{\mu}\partial^{\mu}+\gamma_{\mu}
A^{\mu})\psi_{a}-\frac{1}{4e^{2}_{c}}F_{\mu\nu}F^{\mu\nu}\right]
\label{action}
\end{equation}
The theory has an internal $SU_L(2)\otimes SU_R(2)$-flavor isospin symmetry; 
the Dirac fields are 
an isodoublet whereas the electromagnetic field is an isosinglet.
It is well known that in $1+1$ dimensions there is no spontaneous breakdown 
of continuous internal symmetries, unless there are anomalies or the Higgs 
phenomenon occurs. Neither mechanism is possible in the 2-flavor Schwinger 
model for the $SU_L(2)\otimes SU_R(2)$-symmetry: isovector currents do not 
develop anomalies and there are no gauge 
fields coupled to the isospin currents. The particles belong then to 
isospin multiplets. For what concerns the $U(1)$ gauge symmetry there is an Higgs 
phenomenon~\cite{cjs}. 

The action is invariant under the symmetry 
\beq
SU_{L}(2)\otimes SU_{R}(2)\otimes U_{V}(1) \otimes U_{A}(1) \nonumber
\eeq
The group generators act on the fermion isodoublet to give
\begin{eqnarray}
SU_{L}(2) &:& \psi_{a}(x)\longrightarrow (
e^{i\theta_{\alpha}\frac{\sigma^{\alpha}}{2}P_{L}})_{ab}\ \psi_{b}(x)\   , \  
\overline{\psi_{a}}(x)\longrightarrow \overline{\psi_{b}}\ (x)
(e^{-i\theta_{\alpha}\frac{\sigma^{\alpha}}{2}P_{R}})_{ba}\label{s1} \\
SU_{R}(2) &:& \psi_{a}(x)\longrightarrow 
(e^{i\theta_{\alpha}\frac{\sigma^{\alpha}}{2}P_{R}})_{ab}\ \psi_{b}(x)\   ,\   
\overline{\psi_{a}}(x)\longrightarrow \overline{\psi_{b}}(x)\ 
(e^{-i\theta_{\alpha}\frac{\sigma^{\alpha}}{2}P_{L}})_{ba} \\
U_{V}(1) &:& \psi_{a}(x)\ \longrightarrow 
(e^{i\theta(x){\bf 1}})_{ab}\ \psi_{b}(x)\  ,\  
\psi_{a}^{\dagger}(x)\longrightarrow \psi_{b}^{\dagger}(x)\ 
(e^{-i\theta(x){\bf 1}})_{ba} \\
U_{A}(1) &:& \psi_{a}(x)\longrightarrow (e^{i\alpha 
\gamma_{5}{\bf 1}})_{ab}\ \psi_{b}(x)\  ,\  
\psi_{a}^{\dagger}(x)\longrightarrow \psi_{b}^{\dagger}(x)\ 
(e^{-i\alpha \gamma_{5}{\bf 1}})_{ba}
\quad ,
\label{s4} 
\end{eqnarray}
where $\sigma^{\alpha}$ are the Pauli matrices, $\theta_{\alpha}$, $\theta(x)$ and $\alpha$ are real coefficients and
\beq
P_{L}=\frac{1}{2}(1-\gamma_{5})\ ,\ P_{R}=\frac{1}{2}(1+\gamma_{5})\quad .
\eeq
At the classical level the symmetries (\ref{s1}$-$\ref{s4}) lead to 
conservation laws for the isovector, vector and axial currents
\begin{eqnarray}
j_{\alpha}^{\mu}(x)_{R}&=&\overline{\psi}_{a}(x)\gamma^{\mu}P_{R}
(\frac{\sigma_{\alpha}}{2})_{ab}\psi_{b}(x)
\label{ca}\\
j_{\alpha}^{\mu}(x)_{L}&=&\overline{\psi}_{a}(x)\gamma^{\mu}P_{L}
(\frac{\sigma_{\alpha}}{2})_{ab}\psi_{b}(x)
\label{cb}\\
j^{\mu}(x)&=&\overline{\psi}_{a}(x)\gamma^{\mu}{\bf 1}_{ab}\psi_{b}(x)
\label{cc}\\
j^{\mu}_{5}(x)&=&\overline{\psi}_{a}(x)\gamma^{\mu}\gamma ^{5}
{\bf 1}_{ab}\psi_{b}(x)
\end{eqnarray}
It is well known that at the quantum level the vector and axial currents 
cannot be simultaneously conserved, 
due to the anomaly phenomenon \cite{abj}. If the regularization is gauge 
invariant, so that 
the vector current is conserved, then the axial current acquires the 
anomaly which breaks the $U_{A}(1)$-symmetry 
\beq
\partial_{\mu} j_5^{\mu}(x)=2\frac{e_{c}^{2}}{2\pi}\epsilon_{\mu \nu}
F^{\mu \nu}(x)
\label{anomaly}
\eeq
The isoscalar and isovector chiral condensates are zero due to the 
Coleman theorem \cite{colemant}; in fact, they would break not only the 
$U_A(1)$ symmetry 
of the action, but also the continuum internal symmetry 
$SU_L(2)\otimes SU_R(2)$ down to $SU_V(2)$. 
There is, however,  a $SU_L(2)\otimes SU_R(2)$ invariant operator, 
which is non-invariant under the $U_A(1)$-symmetry;
it can acquire a non-vanishing v.e.v without violating 
Coleman's theorem and consequently 
may be regarded as a good order parameter for the 
$U_A(1)$-breaking.
Its expectation value is given by~\cite{gatt,hoso2}
\begin{equation} 
<F>\equiv
<\overline{\psi}_{L}^{(2)}\overline{\psi}_{L}^{(1)}
\psi_{R}^{(1)}\psi_{R}^{(2)}>=(\frac{e^{\gamma}}{4\pi})^{2} 
\frac{2}{\pi} e_{c}^{2}\quad .
\label{chico}
\end{equation}
It describes a process in which two right movers are anihilated
and two left movers are created. 
Note that $F$, being quadrilinear in the fields,
is actually invariant under chiral rotations of $\pi/2$, namely under the 
discrete axial symmetry
\begin{equation}
\psi_a(x)\to \gamma^5\psi_a(x)\quad\bar\psi_a(x)\to-\bar\psi_a(x)\gamma_5\ \ .
\label{disax}
\end{equation}
As a consequence, this part of the chiral symmetry group is not broken by 
the non-vanishing v.e.v. of $F$ (\ref{chico}).

As we shall see in section 6, the lattice theory faithfully reproduces the 
pattern of symmetry breaking of the continuum theory; this happens even if 
on the lattice the $SU(2)$-flavor symmetry is not protected 
by the Coleman theorem. The isoscalar and isovector chiral 
condensates are zero also on the lattice,
whereas the operator $F$ acquires a non-vanishing v.e.v. due to the 
coupling of left and right movers induced by the gauge field.
The continuous axial symmetry is broken explicitly by 
the staggered fermion, but the discrete axial symmetry (\ref{disax}) remains.

The action (\ref{action}) may be presented 
in usual abelian bosonized form \cite{coleman}. Setting
\beq
:\overline{\psi}_{a}\gamma^{\mu}\psi_{a}:=\frac{1}{\sqrt{\pi}}\epsilon^{\mu 
\nu}\partial_{\nu}\Phi_{a}\  ,\  a=1,2\quad ,
\eeq
the electric charge density and the action read
\beq
j_{0}=:\psi^{\dagger}_{1}\psi_{1}+\psi^{\dagger}_{2}\psi_{2}:=
\frac{1}{\sqrt{\pi}}\partial_{x}(\Phi_{1}+\Phi_{2})
\label{chde}
\eeq
\beq
S=\int d^{2}x \left[\frac{1}{2}\partial_{\mu}\Phi_{1}\partial^{\mu}\Phi_{1}
+\frac{1}{2}\partial_{\mu}\Phi_{2}\partial^{\mu}\Phi_{2}-
\frac{e_{c}^{2}}{2\pi}(\Phi_{1}+\Phi_{2})^{2}\right] \quad .
\eeq
By changing the variables to
\begin{eqnarray}
\Phi_{+}&=&\frac{1}{\sqrt{2}}(\Phi_{1}+\Phi_{2})\\
\Phi_{-}&=&\frac{1}{\sqrt{2}}(\Phi_{1}-\Phi_{2})\quad ,
\end{eqnarray}
one has
\beq
S=\int d^{2}x \left(\frac{1}{2}\partial_{\mu}\Phi_{+}
\partial^{\mu}\Phi_{+}+\frac{1}{2}\partial_{\mu}\Phi_{-}
\partial^{\mu}\Phi_{-}-
\frac{e_{c}^{2}}{\pi}\Phi_{+}^{2}\right)\quad .
\label{boa}
\eeq

The theory describes two scalar fields, one massive and one massless. 
$\Phi_{+}$ is an isosinglet as 
evidenced from Eq.(\ref{chde}); its mass $m_{S}=\sqrt{\frac{2}{\pi}}e_{c}$ 
comes from the anomaly Eq.(\ref{anomaly})~\cite{cjs}. Local electric charge 
conservation is 
spontaneously broken, but no Goldstone boson appears because the 
Goldstone mode may be gauged away.
$\Phi_{-}$ represents an isotriplet; it has rather involved nonlinear 
transformation properties under a general 
isospin transformation. All three isospin currents can be written in 
terms of $\Phi_{-}$ but only the third component has a simple 
representation in terms of $\Phi_-$; namely
\begin{eqnarray}
j_{\mu} ^{3} (x) = : \overline{\psi} _{a} (x) \gamma_{\mu} 
(\frac{\sigma^{3}}{2})_{ab} \psi_{b} (x):=\quad \quad \quad 
\quad \quad \nonumber\\
   = \frac{1}{2}:\overline{\psi}_{1}(x)\gamma_{\mu}\psi_{1}(x)-
\overline{\psi}_{2}(x)\gamma_{\mu}\psi_{2}(x):=(2\pi)^{\frac{1}{2}}
\epsilon^{\mu \nu}
\partial_{\nu}\Phi_{-}\quad .
\end{eqnarray}
The other two isospin currents $j_{\mu}^{1}(x)$ and $j_{\mu}^{2}(x)$ 
are complicated nonlinear 
and nonlocal functions of $\Phi_{-}$ \cite{coleman}; 
a more symmetrical treatment of the bosonized form of the isotriplet currents is 
available within the framework of nobn abelian bosonization~\cite{witten}. 
For the multiflavor Schwinger model this approach has been carried out in~\cite{gepner}, 
providing results in agreement with~\cite{coleman}.

The excitations are most conveniently classified in terms of the 
quantum numbers of $P$-parity and $G$-parity; 
$G$-parity is related to the charge conjugation $C$ by
\beq
G=e^{i\pi \frac{\sigma^{2}}{2}} C\quad .
\eeq
$\Phi_{-}$ is a $G$-even pseudoscalar, while $\Phi_{+}$ is a $G$-odd pseudoscalar
\begin{eqnarray}
\Phi_{-}\  &:&\  I^{PG}=1^{-+}\\
 \Phi_{+}\  &:&\  I^{PG}=0^{--}\ \ .
\end{eqnarray}
The massive meson $\Phi_{+}$ is stable by $G$ conservation since the action 
(\ref{boa}) is invariant under $\Phi_{+}\longrightarrow -\Phi_{+}$.

In the massive $SU(2)$ Schwinger model $-$ when the mass of the fermion 
$m$ is small compared to $e^2$ (strong coupling) $-$ 
Coleman \cite{coleman} showed that $-$ in addition to the triplet $\Phi_-$ 
($I^{PG}=1^{-+}$) $-$ the low-energy spectrum exhibits a singlet 
$I^{PG}=0^{++}$  lying on 
top of the triplet $\Phi_-$. In this limit the gauge theory is 
mapped to a sine-Gordon model and  the low-lying excitations  
are soliton-antisoliton states. When $m\rightarrow 0$, these
soliton-antisoliton states become massless \cite{hara}; in this 
limit, the analysis of the many body wave functions, carried out in 
Ref.\cite{hara}, hints to the existence of  
a whole class of massless states with positive G-parity; 
P-parity however cannot be determined with the procedure developed
in \cite{hara}. 
These are not the only excitations of the model:
way up in mass there is the 
isosinglet $I^{PG}=0^{--}$, ($\Phi_+$), already discussed in 
Ref.~\cite{coleman}. 
The model exhibits also triplets, 
whose mass $-$ of order $m_{S}$ or greater $-$ stays finite~\cite{hara}; 
among the triplets there is a G-even state
\footnote{K. Harada private communication.}.

The Hamiltonian, gauge constraint and non-vanishing (anti-)commutators
of the continuum 2-flavor Schwinger model are
\begin{eqnarray}
H=\int dx&\left[\frac{e^2}{2}E^2(x)+\sum_{a=1}^2
\psi^{\dagger}_a (x)\alpha\left(i\partial_x +eA(x)\right)\psi_a(x)\right]
\label{ham1}\\ 
&\partial_x E(x)\ +\sum_{a=1}^2 \psi^{\dagger}_a
(x)\psi_a (x)\sim 0\label{ga1}\\
&\left[ A(x),E(y)\right]=i\delta(x-y) ~,
 \left\{\psi_a(x),\psi_b^{\dagger}(y)\right\}=\delta_{ab}\delta(x-y)\ \ .
\label{commu1}
\end{eqnarray}
A lattice Hamiltonian, constraint and (anti-) commutators reducing to 
(\ref{ham1},\ref{ga1},\ref{commu1}) 
in the naive continuum limit are 
\begin{eqnarray}
H_{S}=\frac{e^{2}a}{2}\sum_{x=1}^N E_{x}^{2}&-&\frac{it}{2a}\sum_{x=1}^N
\sum_{a=1}^2 \left(\psi_{a,x+1}^{\dag}e^{iA_{x}}\psi_{a,x}
-\psi_{a,x}^{\dag}e^{-iA_{x}}\psi_{a,x+1}\right)\label{hamilton}\nonumber\\
E_{x}-E_{x-1}&+&\psi_{1,x}^{\dag}\psi_{1,x}+\psi_{2,x}^{\dag}\psi_{2,x}-1\sim
0\ ,
\label{gauss}\\
\left[ A_x,E_y\right]=i\delta_{x,y}~&,&
\left\{\psi_{a,x},\psi_{b,y}^{\dagger}\right\}=\delta_{ab}\delta_{xy}\ \ .
\nonumber
\end{eqnarray}
The fermion fields are defined on the sites, $x=1,...,N$, 
the gauge and electric fields, $ A_{ x}$ and
 $E_{x}$,  on the links $[x; x + 1]$, $N$ is an even integer 
and, when $N$ is finite it is convenient to impose periodic boundary conditions.  When $N$ is finite, the continuum limit is the 
2-flavor Schwinger model on a circle \cite{manton}.
The coefficient $t$ of the hopping term in (\ref{hamilton})
plays the role of the lattice light speed. In the naive continuum limit,
$e_L=e_c$ and $t=1$. 

The Hamiltonian and gauge constraint exhibit the discrete symmetries 
\begin{itemize}
\item{}Parity P: 
\begin{equation}
A_{x}\longrightarrow -A_{-x-1},\ E_{x}\longrightarrow -E_{-x-1},\ 
\psi_{a,x}\longrightarrow (-1)^{x}\psi_{a,-x},\ \psi_{a,x}^{\dag}\longrightarrow 
(-1)^{x}\psi_{a,-x}^{\dag}
\label{par}
\end{equation}

\item{}Discrete axial symmetry $\Gamma$: 
\begin{equation}
A_{x}\longrightarrow A_{x+1},\ E_{x}\longrightarrow E_{x+1},\ 
\psi_{a,x}\longrightarrow \psi_{a,x+1},\ \psi_{a,x}^{\dag}\longrightarrow 
\psi_{a,x+1}^{\dag}
\label{chir}
\end{equation}

\item{}Charge conjugation C:
\begin{equation}
A_{x}\longrightarrow -A_{x+1},\  E_{x}\longrightarrow -E_{x+1},\ 
\psi_{a,x}\longrightarrow \psi^{\dag}_{a,x+1},\ \psi_{a,x}^{\dag}\longrightarrow 
\psi_{a,x+1}
\label{char}
\end{equation}

\item{}G-parity:
\begin{eqnarray}
A_{x}\longrightarrow -A_{x+1},\  E_{x}\longrightarrow -E_{x+1}\nonumber\\
\psi_{1,x}\longrightarrow \psi^{\dag}_{2,x+1},\ \psi_{1,x}^{\dag}
\longrightarrow \psi_{2,x+1}\\
\psi_{2,x}\longrightarrow -\psi^{\dag}_{1,x+1},\ \psi_{2,x}^{\dag}
\longrightarrow -\psi_{1,x+1}\ .\nonumber
\end{eqnarray}
\end{itemize}

The lattice 2-flavor Schwinger model is equivalent to a one 
dimensional quantum Coulomb gas on the lattice with two kinds of particles. To see this one can fix the 
gauge, $A_{x} = A$ (Coulomb gauge). Eliminating the non-constant electric 
field and using the gauge constraint, one obtains the effective Hamiltonian
\begin{eqnarray}
H_{S}&=&H_u+H_p
\equiv\left[\frac{e^{2}_{L}}{2 N}E^{2}+\frac{e^{2}_{L}a}{2}
\sum_{x,y}\rho(x) V(x-y)\rho(y)\right]+\nonumber\\
&+&\left[
-\frac{it}{2a}\sum_{x}\sum_{a=1}^{2}(\psi_{a,x+1}^{\dag}e^{iA}\psi_{a,x}-\psi_{a,x}^{\dag}e^{-iA}
\psi_{a,x+1})\right]\ ,
\label{hs}
\end{eqnarray}
where the charge density is
\begin{equation}
\rho(x)=\psi^{\dag}_{1,x}\psi_{1,x}+\psi^{\dag}_{2,x}\psi_{2,x}-1\ \ ,
\label{cd1}
\end{equation}
and the potential
\begin{equation}
V(x-y)=\frac{1}{N}
\sum^{N-1}_{n=1} e^{i 2\pi n (x-y)/N}\frac{1}{4\sin^2\frac{\pi n}{N}}
\end{equation}
is the Fourier transform of the inverse laplacian on the lattice
for non zero momentum.
The constant 
modes of the gauge field decouple in the thermodynamic limit 
$ N \longrightarrow \infty $.

\section{The strong coupling limit and the 
antiferromagnetic Heisenberg Hamiltonian}
In a previous paper \cite{noi2} we showed that the low-lying spectrum of the 
2-flavor lattice Schwinger model in the strong coupling limit 
is equivalent to the spectrum  of the antiferromagnetic Heisenberg model. 
There we also showed that the mass of the massive excitations can be 
computed in terms of correlators of the Heisenberg model.
It is our purpose in this section to further explore this equivalence 
and to set up the formalism needed in what follows. 

In the thermodynamic limit the Schwinger Hamiltonian (\ref{hs}), 
rescaled by the
 factor ${e_{L}^{2}a}/{2}$, reads 
 \begin{equation}
H=H_{0}+\epsilon H_{h}
\end{equation}
with
\begin{eqnarray}
H_{0}&=&\sum_{x>y}\left[\frac{(x-y)^{2}}{N}-(x-y)\right]\rho(x)\rho(y)\quad ,
\label{hu}\\
H_{h}&=&-i(R-L)
\label{hp}
\end{eqnarray}
and $\epsilon=t/e_{L}^{2}a^{2}$.
In Eq.(\ref{hp}) the right $R$ and left $L$ hopping operators are defined ($L=R^{\dagger}$) as
\begin{equation}
R=\sum_{x=1}^{N}R_{x}=\sum_{x=1}^{N}\sum_{a=1}^{2}R_{x}^{(a)}= 
\sum_{x=1}^{N} \sum_{a=1}^2 \psi_{a,x+1}^{\dag}e^{iA}\psi_{a,x}\quad .
\end{equation}
On a periodic chain the commutation relation 
\begin{equation}
[R,L]=0\quad 
\end{equation}
is satisfied.

We shall consider the strong coupling perturbative expansion 
where the Coulomb 
Hamiltonian (\ref{hu}) is the unperturbed Hamiltonian and the hopping 
Hamiltonian 
(\ref{hp}) the perturbation.
Due to Eq.(\ref{cd1}) every configuration with one particle per site 
has zero energy, so that the ground state of the Coulomb 
Hamiltonian (\ref{hu}) is $2^N$ times degenerate. The degeneracy of 
the ground state can be removed only at the second perturbative order 
since the first order is trivially zero. 

At the second order the 
lattice gauge theory is effectively described by the antiferromagnetic 
Heisenberg Hamiltonian. 
The vacuum energy $-$ at order $\epsilon^2$ $-$ reads  
\begin{equation}
E^{(2)}_{0}=<H_{h}^{\dagger}\frac{\Pi}{E_{0}^{(0)}-H_{0}}H_{h}>
\label{secorder}
\end{equation}
where the expectation values are defined on the degenerate subspace and 
$\Pi$ is the operator projecting on a set orthogonal to the states with one 
particle per site. 
Due to the vanishing of the charge density on the ground states of $H_{0}$, 
the commutator
\begin{equation}
[H_0, H_h]=H_h
\label{comm1}
\end{equation}
holds on any linear combination of the degenerate ground states. 
Consequently, from Eq.(\ref{secorder}) one finds
\begin{equation}
E^{(2)}_{0}=-2<RL>\quad .
\label{secorder2}
\end{equation}
On the ground state the combination $R L$ can be written in terms of the 
Heisenberg Hamiltonian.
By introducing the Schwinger spin operators
\begin{equation}
\vec{S}_{x}=\psi_{a,x}^{\dag}\frac{\vec{\sigma}_{ab}}{2}\psi_{b,x}
\end{equation}
the Heisenberg Hamiltonian $H_{J}$ reads
\begin{eqnarray}
H_{J}&=&\sum_{x=1}^{N}\left(\vec{S}_x
\cdot \vec{S}_{x+1}-\frac{1}{4}\right)=\nonumber\\ 
&=&\sum_{x=1}^{N} \left( -\frac{1}{2} 
L_{x}R_{x}-\frac{1}{4}\rho(x) \rho(x+1) \right)
\end{eqnarray}
and, on the degenerate subspace, one has 
\begin{equation}
<H_{J}>=\left<\sum_{x=1}^{N}\left(\vec{S}_x\cdot \vec{S}_{x+1}-\frac{1}{4}
\right)\right>
=\left<\sum_{x=1}^{N}\left(-\frac{1}{2}L_{x}R_{x}\right)\right>\quad .
\label{mainequation}
\end{equation}
Taking into account that products of $L_x$ and $R_y$ at different points 
have vanishing expectation values on the ground states,
and using Eq.(\ref{mainequation}), Eq.(\ref{secorder2}) reads
\begin{equation}
E^{(2)}_{0}=4<H_{J}>\quad .
\label{secorder3}
\end{equation}
The ground state of $H_{J}$ singles out the correct vacuum, on which to 
perform the perturbative expansion. 
In one dimension $H_{J}$ is exactly diagonalizable \cite{bethe, mattis}. 
In the spin model a flavor 1 particle on a site 
can be represented by a spin up, a flavor 2 particle by a spin down. 
The spectrum of $H_{J}$ exhibits $2^N$ eigenstates; among these, the spin 
singlet with lowest energy is the non degenerate ground state $|g.s.>$. 

We shall construct the strong coupling perturbation theory of the 
2-flavor Schwinger model using $|g.s.>$ as the unperturbed ground state. 
$|g.s.>$ is invariant under translations by one lattice site, which amounts
to invariance under discrete chiral transformations. As a consequence, at variance with 
the 1-flavor model~\cite{noi}, chiral symmetry cannot be 
spontaneously broken even in the infinite coupling limit. 
 
$|g.s.>$ has zero charge density on each site and zero electric flux 
on each link 
\begin{equation}
\rho(x)|g.s.>=0\quad ,\quad E_{x}|g.s.>=0 \quad \quad (x=1,...,N)\quad .
\label{keyequation}
\end{equation}
$|g.s.>$ is a linear combination of all the possible states with 
$\frac{N}{2}$ spins up and $\frac{N}{2}$ spins down. The 
coefficients are not explicitly known for general $N$. In the appendix A
we shall exhibit $|g.s.>$ explicitly
for finite size systems of 4, 6 and 8 sites. 
The Heisenberg energy of $|g.s.>$ is known exactly and, in the 
thermodynamic limit, is \cite{hulten,faddeev}
\begin{equation}
H_{J}|g.s.>=(-N\  \ln\ 2)|g.s.>\quad .
\label{mainenergy}
\end{equation}
Eq.(\ref{mainenergy}) provides the second order correction 
Eq.(\ref{secorder3}) to the vacuum energy, $E_{g.s.}^{(2)}=-4 N\ln 2$.

There exist two kinds of excitations created from 
$|g.s.>$; one kind involves only spin flipping and has lower energy 
since no electric flux is created, the other involves fermion 
transport besides spin flipping and thus has a higher energy. For the 
latter excitations the energy is proportional to the coupling times 
the length of the electric flux: the lowest energy is achieved when the 
fermion is transported by one lattice spacing. 
Of course only the excitations of the first kind can be mapped into 
states of the Heisenberg model. 

In~\cite{faddeev} the antiferromagnetic Heisenberg model excitations 
have been classified. There it was shown that any excitation may be 
regarded as 
the scattering state of quasiparticles of spin-$\frac{1}{2}$: every 
physical state contains an even number of quasiparticles and 
the spectrum exhibits only integer spin states. 
The two simplest excitations of lowest energy in the thermodynamic 
limit are a triplet and a singlet \cite{faddeev}; they 
have a dispersion relation depending on the momenta of the two 
quasiparticles. 
For vanishing total momentum (relative to the ground state 
momentum $P_{g.s.} =0$ for $\frac{N}{2}$ even, 
$P_{g.s.} =\pi$ for $\frac{N}{2}$ odd) in the thermodynamic limit
they are degenerate with the ground state. 

In the appendix A we show that even for finite size systems, the excited
states can be grouped in families corresponding to the classification
given in~\cite{faddeev}. We explicitly exhibit all the energy 
eigenstates for $N=4$ and $N=6$. The lowest lying are a triplet and a 
singlet, respectively; they have a well defined relative 
(to the ground state) $P$-parity and $G$-parity
$-$ $1^{-+}$ for the triplet and $0^{++}$ for the singlet.
Since they share the same quantum numbers
these states can be identified, in the limit of vanishing fermion mass, 
with the soliton-antisoliton excitations 
found by Coleman in his analysis of the 2-flavor Schwinger model.
A related analysis about the parity of the lowest lying states 
in finite size Heisenberg chains, has been given in~\cite{eggert}. 

Moreover in \cite{faddeev} a whole class $-$ ${\cal M}_{AF}$ $-$ of 
gapless excitations at zero momentum was singled out 
in the thermodynamic limit; these states are eigenstates of the total 
momentum and consequently have positive G-parity at 
zero momentum. The low-lying states of the Schwinger model also contain 
\cite{hara} many massless excitations with positive G-parity; they are 
identified \cite{noi2} with the excitations belonging to 
${\cal M}_{AF}$. 
The mass of these states in the Schwinger model can be obtained
from the differences between the excitation
energies at zero momentum 
and the ground state energy. 
The energies of the states $|ex.>$ belonging to the class
${\cal M}_{AF}$ have the same 
perturbative expansion of the ground state. 
Consequently, the states $|ex.>$ 
at zero momentum up to the second order in the strong coupling 
expansion have the same energy 
of the ground state (\ref{secorder}), 
$E^{(2)}_{ex}=-4 N\ln 2$.  To this order the
mass gap is zero. 
Higher order corrections may give a mass gap.

\section{The meson masses}

In this section we determine the masses 
for the states obtained by fermion transport
of one site on the Heisenberg model ground state.
Our analysis shows that besides the $G$-odd pseudoscalar isosinglet
$0^{--}$ with mass $m_S=e\sqrt{2/\pi}$, there are also a 
$G$-even pseudoscalar isosinglet $0^{++}$ and isotriplet $1^{-+}$
and a $G$-odd scalar isotriplet $1^{+-}$ with masses of the order of 
$m_S$ or greater.  
The quantum numbers are relative to those of the ground state
$I_{g.s.}^{PG}=0^{++}$ for $N/2$ even $I_{g.s.}^{PG}=0^{--}$ for $N/2$
odd.

Two states can be created using the spatial
component of the vector $j^{1}(x)$ Eq.(\ref{cc}) and 
isovector $j_{\alpha}^{1}(x)$ Eqs.(\ref{ca},\ref{cb}) 
Schwinger model currents. They are the G-odd pseudoscalar 
isosinglet $I^{PG}=0^{--}$ and the
G-even pseudoscalar isotriplet $I^{PG}=1^{-+}$.  The lattice operators
with the correct quantum numbers creating these states at zero momentum, 
when acting on $|g.s.>$, read 
\begin{eqnarray}
S&=&R+L=\sum_{x=1}^{N}j^{1}(x)\\
T_{+}&=&(T_{-})^{\dagger}=R^{(12)}+L^{(12)}=\sum_{x=1}^{N}j_{+}^{1}(x)
\label{ta}\\
T_{0}&=&\frac{1}{\sqrt{2}}(R^{(11)}+L^{(11)}-R^{(22)}-L^{(22)})=
\sum_{x=1}^{N}j_{3}^{1}(x) \quad .
\label{t0}
\end{eqnarray}
$R^{(ab)}$ and $L^{(ab)}$ in (\ref{ta},\ref{t0}) are the 
right and left flavor changing 
hopping operators ($L^{(ab)}=(R^{(ab)})^{\dagger}$)
$$
R^{(ab)}=\sum_{x=1}^{N}\psi_{a,x+1}^{\dagger}e^{iA}\psi_{b,x}\quad .
$$
The states are given by
\begin{eqnarray}
|S>&=&|0^{--}>=S|g.s.>\\
|T_{\pm}>&=&|1^{-+},\pm1>=T_{\pm}|g.s.>
\label{2tpm}\\
|T_{0}>&=&|1^{-+},0>=T_{0}|g.s.>\ .
\label{to}
\end{eqnarray}
They are normalized as
\begin{eqnarray}
<S|S>&=&<g.s.|S^{\dagger}S|g.s.>=-4<g.s.|H_{J}|g.s.>=4N \ln 2
\label{no1}\\
<T_{+}|T_{+}>&=&\frac{2}{3}(N+<g.s.|H_{J}|g.s.>)=\frac{2}{3}N(1-\ln 2)
\label{no2}
\end{eqnarray}
and 
\begin{equation}
<T_{0}|T_{0}>=<T_{-}|T_{-}>=<T_{+}|T_{+}> \ .
\label{no3}
\end{equation}
In Eqs.(\ref{no1},\ref{no2},\ref{no3}) $<g.s.|g.s.>=1$.

The isosinglet energy, up to the second order in the strong coupling 
expansion, is
$
E_{S}=E_{S}^{(0)}+\epsilon^{2}E_{S}^{(2)}
$ with
\begin{eqnarray}
E_{S}^{(0)}&=&\frac{<S|H_{0}|S>}{<S|S>}=1\quad ,\\
E_{S}^{(2)}&=&\frac{<S|H_{h}^{\dag}\Lambda_{S}H_{h}|S>}{<S|S>}\quad ,
\label{secordex}
\end{eqnarray}
$
\Lambda_{S}=\frac{\Pi_{S}}{E_{S}^{(0)}-H_{0}}
$
and $1-\Pi_{S}$ a projection operator onto $|S>$.
On $|g.s.>$ 
\begin{equation}
[H_{0},(\Pi_{S}H_{h})^n S]=(n+1)(\Pi_{S}H_{h})^n S 
,\quad
(n=0,1,\dots),
\end{equation}
holds; Eq.(\ref{secordex}) may then be written in terms of spin correlators as
\begin{equation}
E_{S}^{(2)}=E_{g.s.}^{(2)}+4-\frac{\sum_{x=1}^{N}
<g.s.|\vec{S}_{x}\cdot \vec{S}_{x+2}-\frac{1}{4}|g.s.>}{<g.s.|H_{J}|g.s.>}
\quad .
\label{senes}
\end{equation}
One immediately recognizes that the excitation spectrum is determined 
once $<g.s.|\vec{S}_{x}\cdot \vec{S}_{x+2}|g.s.>$ is known. 
Equations similar to Eq.(\ref{senes}) may be established also at a 
generic order of the strong coupling expansion.    

At the zeroth perturbative order the pseudoscalar triplet is 
degenerate with the isosinglet $E_{T}^{(0)}=E_{S}^{(0)}=1$. 
Following the same procedure as before one may compute the energy of the 
states (\ref{2tpm}) and 
(\ref{to}) to the second order in the strong coupling expansion. To 
this order, the energy is given by
\begin{eqnarray}
&&E_{T}^{(2)}=E_{g.s.}^{(2)}-\Delta_{DS}(T)-\nonumber\\ 
&&\frac{4<g.s.|H_{J}|g.s.>+5\sum_{x=1}^{N}
<g.s.|\vec{S}_{x}\cdot \vec{S}_{x+2}-\frac{1}{4}|g.s.>}{N+<g.s.|H_{J}|g.s.>}
\label{senet}
\end{eqnarray}
where in terms of the vector operator 
$
\vec{V}=\sum_{x=1}^{N}\vec{S}_{x}\wedge \vec{S}_{x+1}
$, one can write $\Delta_{DS}(T)$ as
\begin{eqnarray}
\Delta_{DS}(T_{\pm})&=&12
\frac{<g.s.|(V_{1})^2|g.s.>+<g.s.|(V_{2})^2|g.s.>}
{N+<g.s.|H_{J}|g.s.>}\\
\Delta_{DS}(T_{0})&=&12
\frac{2<g.s.|(V_{3})^2|g.s.>}{N+<g.s.|H_{J}|g.s.>}\ .
\end{eqnarray}
The v.e.v. of each squared component of $\vec{V}$ on 
the rotationally invariant singlet $|g.s.>$ give the same 
contribution $i.e.$ $\Delta_{DS}(T_{\pm})=\Delta_{DS}(T_{0})$: the 
triplet states (as in the continuum theory) have a degenerate mass gap. 
This is easily verified by direct computation on finite size systems; 
when the size of the system is finite one may also show that $\Delta_{DS}$ 
is of zeroth order in N.

The excitation masses are given by 
$m_{S}=\frac{e_{L}^2 a}{2}(E_{S}-E_{g.s.})$ and 
$m_{T}=\frac{e_{L}^2 a}{2}(E_{T}-E_{g.s.})$. 
Consequently, the ($N$-dependent) ground state energy terms 
appearing in $E_{S}^{(2)}$ and $E_{T}^{(2)}$ cancel and what is left 
are only $N$ independent terms.
This is a good check of our computation, being the mass an intensive quantity.
 
In principle one should expect also excitations created acting 
on $|g.s.>$ with the chiral currents, in analogy with the one flavor 
Schwinger model where, as shown in Ref.~\cite{kogut}, the chiral current 
creates a two-meson bound state. The chiral currents operators for the two 
flavor Schwinger model are given by
\begin{eqnarray}
j^{5}(x)&=&\overline{\psi}(x)\gamma^{5}\psi(x)\\
j^{5}_{\alpha}(x)&=&\overline{\psi}_{a}(x)\gamma^{5}
(\frac{\sigma}{2})_{ab}\psi_{b}(x)\quad .
\end{eqnarray}
The corresponding lattice operators at zero momentum are 
\begin{eqnarray}
S^{5}&=R-L=\sum_{x=1}^{N}j^{5}(x)
\label{s5}\\
T_{+}^{5}&=(T_{-}^{5})^{\dagger}=R^{(12)}-L^{(12)}=
\sum_{x=1}^{N}j_{+}^{5}(x)
\label{ta2}\\
T_{0}^{5}&=\frac{1}{\sqrt{2}}(R^{(11)}-L^{(11)}-R^{(22)}+L^{(22)})=
\sum_{x=1}^{N}j_{3}^{5}(x)\quad .
\label{t02}\\
\end{eqnarray}
The states created by (\ref{s5},\ref{ta2},\ref{t02}) when acting on $|g.s.>$,
are
\begin{eqnarray}
|S^{5}>&=&|0^{++}>=S^{5}|g.s.>\\
|T_{\pm}^{5}>&=&|1^{+-},\pm1>=T_{\pm}^{5}|g.s.>
\label{tpm2}\\
|T_{0}^{5}>&=&|1^{+-},0>=T_{0}^{5}|g.s.>\quad .
\label{to2}
\end{eqnarray}
They are normalized as
\begin{eqnarray}
<S^{5}|S^{5}>&=&<g.s.|S^{5\dagger}S^{5}|g.s.>=-4<g.s.|H_{J}|g.s.>=4N\log 2\\
<T^{5}_{+}|T^{5}_{+}>&=&\frac{2}{3}(N+<g.s.|H_{J}|g.s.>)=\frac{2}{3}
N(1-\log 2)
\end{eqnarray}
and 
\begin{equation}
<T^{5}_{0}|T^{5}_{0}>=<T^{5}_{-}|T^{5}_{-}>=<T^{5}_{+}|T^{5}_{+}>\quad .
\end{equation}

Following the computational scheme used to study $|S>$ and $|T>$, 
one finds for the state $|S^{5}>$
\begin{eqnarray} 
E_{S^5}^{(0)}&=&1\\
E_{S^5}^{(2)}&=&E_{g.s.}^{(2)}+12-3\frac{\sum_{x=1}^{N}
<g.s.|\vec{S}_{x}\cdot \vec{S}_{x+2}-\frac{1}{4}|g.s.>}{<g.s.|H_{J}|g.s.>}\quad .
\label{senes5}
\end{eqnarray}
For the triplet $|T^{5}>$ one gets
\begin{eqnarray}
E_{T^5}^{(0)}&=&1\\
E_{T^5}^{(2)}&=&E_{g.s.}^{(2)}+\frac{\sum_{x=1}^{N}<g.s.|\vec{S}_{x}
\cdot\vec{S}_{x+2}-
\frac{1}{4}|g.s.>-4<g.s.|H_{J}|g.s.>}{N+<g.s.|H_{J}|g.s.>}\quad .
\label{senet5}
\end{eqnarray} 

\section{The correlators $\sum_{x}<g.s.|\vec{S}_{x}\cdot 
\vec{S}_{x+2}|g.s.>$ and $<g.s.|\vec{V}^2|g.s.>$}

In this section we compute the spin-spin 
correlators needed to evaluate the second order energies of the 
isosinglet and isotriplets derived in section 4.
The explicit computation of spin-spin correlation functions is far from 
being trivial since 
$G(r)\equiv <g.s.|\vec{S}_{0}\cdot \vec{S}_{r}|g.s.>$ is not known for 
arbitrary lattice separations $r$.
For $r=2$ it was computed by M. Takahashi \cite{taka} 
in his perturbative analysis of the half filled Hubbard model in 
one dimension. For $r>2$ no exact numerical 
values of $G(r)$ are known. In \cite{korepin} were given two 
representations of $G(r)$, while in \cite{luky, coraff} the exact asymptotic 
($r\rightarrow \infty$) expression of $G(r)$ was derived. 

In order to explicitly compute the second order energies Eq.(\ref{senes}) and 
Eq.(\ref{senet}) one has to evaluate the correlation function
\begin{equation}
G(2)=\frac{1}{N}\sum_{x=1}^{N}<g.s.|\vec{S}_{x}\cdot \vec{S}_{x+2}|g.s.>
\label{corrd2}
\end{equation}
which has been exactly computed in~\cite{taka} and is given by
\begin{equation}
G(2)=\frac{1}{4}(1-16\ln 2+9\zeta (3))=0.1820\quad . 
\label{corrd2n}
\end{equation}
In the appendix B we shall show how the knowledge of this correlator 
allows one to compute explicitly the first three ``emptiness formation 
probabilities'', used in Ref.\cite{korepin} in the study of the Heisenberg
chain correlators, $G(r)$.

In order to compute the mass of the isotriplet one has to evaluate 
also the correlation 
functions appearing in Eq.(\ref{senet}); namely 
$<g.s.|\vec{V}\cdot \vec{V}|g.s.>$, where $\vec V=\sum_x S_x\wedge S_{x+1}$.
The explicit expression of this function is not known.
For its evaluation it is most useful to rearrange this correlator as 
\begin{eqnarray}
& &<g.s.|\vec{V}\cdot \vec{V}|g.s.>=\sum_{x,y=1}^{N}
(<g.s.|(\vec{S}_{x}\cdot \vec{S}_{y})(\vec{S}_{x+1}
\cdot \vec{S}_{y+1})|g.s.>\nonumber\\
&-&<g.s.|(\vec{S}_{x}\cdot \vec{S}_{y+1})(\vec{S}_{x+1}
\cdot \vec{S}_{y})|g.s.>)
-\sum_{x=1}^{N}<g.s.|\vec{S}_{x}\cdot \vec{S}_{x+1}|g.s.>\quad \quad .
\label{vv}
\end{eqnarray}
It is possible to extract a numerical value from Eq.(\ref{vv}) only 
within the random phase approximation \cite{taka, zu}. 
For this purpose it is first convenient 
to rewrite the unconstrained sum over the sites $x$ and $y$ as a sum 
where all the four spins involved in the v.e.v.'s lie on different sites,
\begin{eqnarray}
<g.s.|\vec{V}\cdot \vec{V}|g.s.>&=&\sum_{
\stackrel{y\neq x}{ y\neq x\pm 1}}(<g.s.|(\vec{S}_{x}\cdot 
\vec{S}_{y})(\vec{S}_{x+1}\cdot \vec{S}_{y+1})|g.s.>
\nonumber\\
&-&<g.s.|(\vec{S}_{x}\cdot \vec{S}_{y+1})(\vec{S}_{x+1}\cdot 
\vec{S}_{y})|g.s.>)+\frac{3}{8}N\nonumber\\
&-&\frac{1}{2}\sum_{x=1}^{N}<g.s.|\vec{S}_{x}\cdot \vec{S}_{x+1}|g.s.>
-\sum_{x=1}^{N}
<g.s.|\vec{S}_{x}\cdot \vec{S}_{x+2}|g.s.>
\label{vvv}
\end{eqnarray}
and then factorize the four spin operators in Eq.(\ref{vvv}) as 
\begin{eqnarray}
<g.s.|\vec{V}\cdot \vec{V}|g.s.>&\simeq&N\sum_{r=2}^{\infty}
(<g.s.|\vec{S}_{0}\cdot \vec{S}_{r}|g.s.>^2 -<g.s.|\vec{S}_{0}\cdot 
\vec{S}_{r+1}|g.s.>
<g.s.|\vec{S}_{1}\cdot \vec{S}_{r}|g.s.>)\nonumber\\
&+&\frac{3}{8}N-\frac{1}{2}\sum_{x=1}^{N}<g.s.|\vec{S}_{x}\cdot 
\vec{S}_{x+1}|g.s.>-\sum_{x=1}^{N}<g.s.|\vec{S}_{x}\cdot 
\vec{S}_{x+2}|g.s.>\quad .
\label{rpavv}
\end{eqnarray}
Of course, Eq.(\ref{rpavv}) provides an answer larger than the exact result; 
terms such as $<(\ldots)(\ldots)>$ contain negative contributions 
which are eliminated once one factorizes them in the form $<(...)><(...)>$. 
This is easily checked also by direct computation on finite size systems. 

The spin-spin correlation functions $G(r)$ are exactly known for $r=1,2$. 
For the spin-spin correlation functions $G(r)$ up to a distance of 
$r=30$ the results are reported in table (\ref{30c})  \cite{luky,camp}. 
\begin{table}[htbp]
\begin{center}
\caption{Spin-spin correlation functions}\label{30c}
\vspace{.1in}
\begin{tabular}{|rc|rc|}\hline
$r$   &   $G(r)$ & $r$   &  $G(r)$ \rule{0in}{4ex}\\[2ex] \hline 
 
 1    & -0.4431              & 16    &   0.0305 \rule{0in}{4ex}\\
  
 2    & 0.1821               & 17    &  -0.0296 \\

 3    & -0.1510              & 18    &   0.0274 \\
 
 4    & 0.1038               & 19    &  -0.0267 \\
 
 5    & -0.0925              & 20    &   0.0249 \\
 
 6    & 0.0731               & 21    &   -0.0242 \\
 
 7    & -0.0671              & 22    &   0.0228 \\

 8    & 0.0567              & 23    &   -0.0223 \\
 
 9    & -0.0532              & 24    &   0.0211 \\

 10   & 0.0465              & 25    &   -0.0206 \\
 
 11   &-0.0442              & 26    &   0.0196 \\
 
 12   & 0.0395              & 27    &   -0.0193 \\
 
 13   & -0.0379              & 28    &   0.0183 \\
 
 14   & 0.0344              & 29    &   -0.0181 \\
 
 15   & -0.0332              & 30    &   0.0172 \\[2ex] \hline
\end{tabular}
\end{center}
\end{table}

For $r>30$, one may write \cite{luky}

\begin{eqnarray}
G(r)&=&\frac{3}{4}\sqrt{\frac{2}{\pi^3}}\frac{1}{r\sqrt{g(r)}}
[ 1-\frac{3}{16}g(r)^2+\frac{156\zeta(3)-73}{384}g(r)^3+O(g(r)^4)
-\nonumber \\& &\frac{0.4}{2r}((-1)^r +1+O(g(r))+O(\frac{1}{r^2}) ]
\label{l1}
\end{eqnarray}
with $g(r)$ satisfying
\begin{equation}
g(r)=\frac{1}{C(r)}(1+\frac{1}{2}g(r)\ln (g(r)))
\label{l2}
\end{equation}
and 
\begin{equation}
C(r)=\ln(2\sqrt{2\pi}e^{\gamma +1} r)\quad .
\label{l3}
\end{equation}
Eq.(\ref{l2}) may be solved by iteration.
To the lowest order in $\frac{1}{C}$ one finds
\begin{equation}
g(r)\approx \frac{1}{C(r)}-\frac{1}{C(r)^2}\ln C(r)\quad .
\label{lg}
\end{equation}
Inserting (\ref{lg}) in (\ref{l1}) leads to 
\begin{equation}
G(r)\approx \sqrt{2}{\pi^3}\frac{1}{r}\sqrt{C(r)} 
[1+\frac{1}{4C(r)}\ln C(r)]+O(\frac{1}{C(r)^2})\quad .
\label{sslu}
\end{equation}
Putting (\ref{sslu}) in (\ref{rpavv}), one finally gets  
\begin{equation}
<g.s.|\vec{V}\cdot \vec{V}|g.s.>=0.3816N\ .
\label{vvapp}
\end{equation}

Using Eq.(\ref{corrd2n}), the isosinglet mass reads as
\begin{equation}
\frac{m_{S}}{e^2 a}=\frac{1}{2}+1.9509 \quad \epsilon ^2\quad .
\label{ms} 
\end{equation}
For what concerns the isotriplet mass, since the double sum in 
Eq.(\ref{senet}) is given by
\begin{equation}
\Delta_{DS}(T)=8\frac{<g.s.|\vec{V}\cdot \vec{V}|g.s.>}
{N+<g.s.|H_{J}|g.s.>}\quad ,
\end{equation}
using Eq.(\ref{vvapp}), one gets
\begin{equation}
\frac{m_{T}}{e_{L}^2 a}=\frac{1}{2}+0.0972\quad \epsilon ^2\quad .
\label{mtt} 
\end{equation}

The existence of massive isotriplets was already noticed in \cite{hara}, 
and their mass in the continuum theory was numerically computed 
for various values of the fermion mass. 
In particular there is a $G$-parity even isotriplet with mass 
approximately equal to the mass of the isosinglet $0^{--}$.  

The mass of the $|S_{5}>$ isosinglet and the $|T_{5}>$ isotriplet is
\begin{eqnarray}
\frac{m_{S^5}}{e^2a}&=&\frac{1}{2}+5.85\epsilon^2\label{ms5}\\ 
\frac{m_{T^5}}{e^2a}&=&\frac{1}{2}+4.4069\epsilon^2\ . 
\label{mt5}
\end{eqnarray}

Equations (\ref{ms}), (\ref{mtt}), (\ref{ms5} and (\ref{mt5}) 
provide the values of $m_{S}$, $m_{T}$ and 
$m_{T^5}$ for small values of 
$z=\epsilon^2=\frac{t^2}{e_{L}^4a^4}$ up to the second order 
in the strong coupling expansion. 
Whereas (\ref{mtt}) is only approximate (\ref{ms}) and (\ref{mt5}) 
are exact at the second order in the $\epsilon$ expansion.
In section 6 we shall 
extrapolate these masses to the continuum limit using the 
standard technique of the Pad\'e approximants.

\section{The chiral condensate}

In the following we shall first prove that also on the 
lattice either the isoscalar $\left<\bar\psi\psi\right>$ or the isovector 
$\left< \bar\psi \sigma^a\psi\right>$ chiral condensates are 
zero to every order of perturbation theory. 
This should be verified by explicit computation since on the 
lattice the symmetry $SU_{L}(2)\otimes SU_{R}(2)$ 
is already broken by introducing staggered fermions; 
thus, there is no symmetry to prevent the formation of such 
chiral condensates. In the continuum theory, instead, the breaking of the 
$SU_{L}(2)\otimes SU_{R}(2)$ down to $SU_{V}(2)$ is prevented 
by the Coleman theorem \cite{colemant}.

In the staggered fermion formalism the isoscalar condensate is given by
\begin{equation}
\sum_{a=1}^2 \overline{\psi_{a}}(x)\psi_{a}(x) \longrightarrow
 \frac{(-1)^{x}}{2a}(\psi^{\dagger}_{1,x}\psi_{1,x}+
 \psi^{\dagger}_{2,x}\psi_{2,x}-1)\quad ;
\end{equation}
it is obtained  by considering the mass operator
\begin{equation}
M=\frac{1}{2Na}\sum_{x=1}^{N}(-1)^{x}(\psi^{\dagger}_{1,x}\psi_{1,x}+ 
\psi^{\dagger}_{2,x}\psi_{2,x})
\label{mmmm}
\end{equation}
and evaluating its expectation value on the perturbed states 
$|p_{g.s.}>$ generated by applying $H_{h}$ to $|g.s.>$.
To the second order in the strong coupling expansion, one has
\begin{equation}
|p_{g.s.}>=|g.s.>+\epsilon|p_{g.s.}^{1}>+\epsilon^{2}|p_{g.s.}^{2}>+\dots 
\end{equation} 
where
\begin{eqnarray}
|p_{g.s.}^{1}>&=&-H_{h}|g.s.> \\
|p_{g.s.}^{2}>&=&\frac{\Pi_{g.s.}}{2}H_{h}H_{h}|g.s.>\quad .
\end{eqnarray}
To the fourth order, (\ref{mmmm}) is given by
\begin{equation}
\chi_{isos.}=\frac{<p_{g.s.}|M|p_{g.s.}>}{<p_{g.s.}|p_{g.s.}>}=
\frac{<g.s.|M|g.s.>+\epsilon^{2}<p_{g.s.}^{1}|M|p_{g.s.}^{1}>+\epsilon^{4
}<p_{g.s.}^{2}|M|p_{g.s.}^{2}>+\ldots}{<g.s.|g.s.>+
\epsilon^{2}<p_{g.s.}^{1}|p_{g.s.}^{1}>+\epsilon^{4}
<p_{g.s.}^{2}|p_{g.s.}^{2}>+\ldots}\quad .
\label{ciso}
\end{equation}
It is very easy to see that $\chi_{isos.}$ is zero to all orders in the 
strong coupling expansion.
Let us introduce the translation operator
\begin{equation}
\hat{T}=e^{i \hat{p} a}\quad ;
\end{equation}
using 
\begin{eqnarray}
\hat{T}M\hat{T}^{-1}&=&-M
\label{t1}\\
\hat{T}H_{h}\hat{T}^{-1}&=&H_{h}
\label{t2}
\end{eqnarray}
and 
\begin{equation}
\hat{T}|g.s.>=\pm|g.s.>
\label{tpm}
\end{equation}
one gets order by order in the strong coupling expansion in Eq.(\ref{ciso})
\begin{equation}
\chi_{isos.}=-\chi_{isos.}\quad .
\end{equation}
In Eq.(\ref{tpm}) the $+$ appears when $N/2$ is even and the $-$ when 
$N/2$ is odd.

The isovector chiral condensate is given by the expectation value of the operator
\begin{equation}
\vec{\Sigma}=\frac{1}{2Na}\sum_{x=1}^{N}(-1)^x\psi_{a,x}^{\dagger}
\vec{\sigma}_{ab}\psi_{b,x}
\end{equation}
on the perturbed states $|p_{g.s.}>$.
Taking into account that
\begin{equation}
\hat{T}\Sigma \hat{T}^{-1}=-\Sigma
\end{equation}
one gets
\begin{equation}
\chi_{isov.}=-\chi_{isov.}\quad ;
\end{equation}
also the isovector chiral condensate is identically zero. 

In the continuum there is, as evidenced in section 2, only 
a non-vanishing chiral condensate associated to the anomalous breaking 
of the $U_{A}(1)$ symmetry \cite{gatt, hoso2}. Since 
\begin{equation}
\overline{\psi^{a}}_{L}(x)\psi^{a}_{R}(x)=
\overline{\psi^{a}}(x)\frac{1+\gamma_{5}}{2}\psi^{a}(x)\quad ,
\label{chide}
\end{equation}
the pertinent operator is 
$F=\overline{\psi}_{L}^{(2)}\overline{\psi}_{L}^{(1)}\psi_{R}^{(1)}
\psi_{R}^{(2)}$; 
its expectation value has been computed in 
\cite{gatt}\cite{hoso2} and is given by
\begin{equation}
<\overline{\psi}_{L}^{(2)}\overline{\psi}_{L}^{(1)}\psi_{R}^{(1)}
\psi_{R}^{(2)}>=(\frac{e^{\gamma}}{4\pi})^{2} \frac{2}{\pi} e_{c}^{2}=
(\frac{e^{\gamma}}{4\pi})^{2} m_{S}^{2} \quad.
\label{opua}
\end{equation}

On the lattice one has 
\begin{equation}
\overline{\psi^{a}}_{L}(x)\psi^{a}_{R}(x)\longrightarrow \frac{1}{2a} 
\frac{1}{2}(\psi^{\dagger}_{a,x}\psi_{a,x}-
\psi^{\dagger}_{a,x+1}\psi_{a,x+1}+L_{x}^{(a)}-R_{x}^{(a)})\quad .
\label{chidel}
\end{equation}
The factor $1/2a$ is due to the doubling of the lattice spacing 
in the antiferromagnetic bipartite lattice. 
Upon introducing the occupation number operators $n_{x}^{(a)}=
\psi_{a,x}^{\dagger}\psi_{a,x}$, 
the umklapp operator $F$ is represented on the lattice by
\begin{equation}
F=-\frac{1}{16a^2 N}\sum_{x=1}{N}\left\{
(n_x ^{(1)}-n_{x+1} ^{(1)})(n_x ^{(2)}-n_{x+1} ^{(2)})+
(L_{x}^{(1)}-R_{x}^{(1)})(L_{x}^{(2)}-R_{x}^{(2)})\right\}\quad .
\label{fop}
\end{equation}
The strong coupling expansion carried up to the second order in 
$\epsilon=\frac{t}{e_{L}^{2}a^2}$, 
yields 
\begin{equation}
<F>=\frac{<p_{g.s.}|F|p_{g.s.}>}{<p_{g.s.}|p_{g.s.}>}=
\frac{<g.s.|F|g.s.>+\epsilon^{2}<p_{g.s.}^{1}|F|p_{g.s.}^{1}>}{<g.s.|g.s.>+
\epsilon^{2}<p_{g.s.}^{1}|p_{g.s.}^{1}>}
\label{opuad}
\end{equation}
for the lattice chiral condensate. Since 
\begin{eqnarray}
<g.s.|g.s.>&=&1\\
<p_{g.s.}^{1}|p_{g.s.}^{1}>&=&-4<g.s.|H_{J}|g.s.>\quad ,
\end{eqnarray}
and taking into account that
\begin{eqnarray}
<g.s.|F|g.s.>&=&\frac{1}{8a^{2} N}<g.s.|H_{J}|g.s.>\\
<p^{1}_{g.s.}|F|p^{1}_{g.s.}>&=&
\frac{1}{4a^{2}N}(-2<g.s.|( H_{J})^{2}|g.s.>-\frac{5}{3}
<g.s.|H_{J}|g.s.>+\frac{5}{12}N\nonumber\\
&-&\frac{2}{3}\sum_{x=1}^{N}<g.s.|\vec{S}_{x}\cdot \vec{S}_{x+2}-
\frac{1}{4}|g.s.> )\quad ,
\end{eqnarray}
from Eqs.(\ref{mainenergy}) and (\ref{corrd22}), one gets
\begin{equation}
<F>=\frac{1}{a^{2}}(0.0866-0.4043\epsilon ^{2})\quad .
\label{nopuad}
\end{equation}
A nonvanishing value of the lattice chiral condensate 
is due to the coupling $-$ induced by 
the lattice gauge field $-$ between the right and left fermions. 
This is the relic in the lattice
of the $U_{A}(1)$ anomaly in the continuum theory.

\section{Lattice versus continuum}

We now want to compare our lattice results with the exact 
results of the continuum model; to do this, 
one should extrapolate the strong-coupling expansion 
derived under the assumption that the parameter 
$z=\epsilon ^2=\frac{t^2}{e_{L}^4 a^4} \ll 1$ 
to the region in which $z\gg 1$; this corresponds 
to take the continuum limit since $e_{L}^4 a^4 
\longrightarrow 0$ when $z\longrightarrow \infty$. To make the 
extrapolation possible, it is customary to make use 
of Pad\'e approximants, which allow to 
extrapolate a series expansion beyond the convergence radius. Strong-coupling 
perturbation theory improved by Pad\'e approximants should 
then provide results consistent with the continuum 
theory. As we shall see the strong-coupling expansion derived in this paper 
provides accurate estimates of the meson masses, already at the first 
order in powers of $z$. 

Let us now evaluate $m_{S}$ and the lattice light velocity $t$.
We first compute the ratio between the continuum value of the meson mass 
$m_{S}=\sqrt{\frac{2}{\pi}}e_{c}$ and the lattice coupling constant $e_{L}$ 
by equating the lattice chiral condensate, Eq.(\ref{nopuad}), 
to its continuum counterpart Eq.(\ref{opua})
\begin{equation}
\frac{1}{a^{2}}(0.0866-0.4043z)=(\frac{e^{\gamma}}{4\pi})^{2} m_{S}^{2}\ .
\label{chima1}
\end{equation}
Eq.(\ref{chima1}) is true only when Pad\'e approximants are used since, 
as it stands, the left hand side holds only 
for $z\ll 1$, while the right-hand side provides the value of the 
chiral condensate to be obtained when $z=\infty$. 
Using 
\begin{equation}
a=\frac{t^{\frac{1}{2}}}{e_{L}z^{\frac{1}{4}}}\quad ,
\end{equation}
one gets from Eq.(\ref{chima1})
\begin{equation}
(\frac{m_{S}}{e_{L}})^{2}=(\frac{4\pi }{e^{\gamma}})^{2}
\frac{z^{\frac{1}{2}}}{t}(0.0866-0.4043z)\quad .
\label{chima2}
\end{equation}
As in Refs.\cite{kogut, noi}, due to the factor 
$z^{\frac{1}{2}}$, the second power of Eq.(\ref{chima2})
 should be considered in order to 
construct a non diagonal Pad\'e approximant. 
Since the strong coupling expansion has been carried 
out up to second order in $z$, one is allowed 
to construct only the $[0,1]$ Pad\'e approximant for 
the polynomial written in Eq.(\ref{chima2}). One gets
\begin{equation}
(\frac{m_{S}}{e_{L}})^{4}=(\frac{4\pi}{e^{\gamma}})^{4}
\frac{1}{t^2}\frac{0.0074z}{1+9.3371z}\quad ,
\label{chima3}
\end{equation}
and, taking the continuum limit $z\rightarrow \infty$, one finds
\begin{equation}
(\frac{m_{S}}{e_{L}})^4 =(\frac{4\pi}{e^{\gamma}})^{4} 
\frac{0.0008}{t^{2}}\quad .
\label{chima4}
\end{equation}
 
Next we compute the same mass ratio by equating the 
singlet mass gap given in Eq.(\ref{ms}) to its 
continuum counterpart $m_{S}$
\begin{equation}
e_{L}^{2}a(\frac{1}{2}+1.9509z)=m_{S}\quad .
\label{ma1}
\end{equation}
Again, Eq.(\ref{ma1}) is true only when Pad\'e approximants are used. 
Dividing both sides of Eq.(\ref{ma1}) by $e_{L}$ and taking into account that 
\begin{equation}
e_{L}a=\frac{t^{\frac{1}{2}}}{z^{\frac{1}{4}}}
\end{equation}
one gets
\begin{equation}
\frac{m_{S}}{e_{L}}=\frac{t^{\frac{1}{2}}}
{z^{\frac{1}{4}}}(\frac{1}{2}+1.9509z)\quad .
\label{ma2}
\end{equation}
Taking the fourth power and constructing the $[1,0]$ 
Pad\'e approximant for the right hand side of Eq.(\ref{ma2}) one has
\begin{equation}
(\frac{m_{S}}{e_{L}})^{4}=\frac{t^{2}}{z}(\frac{1}{16}+0.9754z)\quad ;
\label{ma3}
\end{equation}
when $z\rightarrow \infty$, Eq.(\ref{ma3}) gives 
\begin{equation}
(\frac{m_{S}}{e_{L}})^{4}=t^2 0.9754\quad .
\label{ma4}
\end{equation}

The numerical value of the hopping parameter $t$, 
determined if one equates Eq.(\ref{ma4}) and Eq.(\ref{chima4}), is
\begin{equation}
t=\frac{4\pi}{e^{\gamma}}0.1692=1.1940
\label{tequ}
\end{equation}
and lies $19\%$ above the exact value.
Putting this value of $t$ in Eq.(\ref{chima4}) or Eq.(\ref{ma4}) one gets
\begin{equation}
\frac{m_{S}}{e_{L}}=1.0969
\end{equation}
which lies $37\%$ above the exact value $\sqrt{\frac{2}{\pi}}$. 
It is comforting to see that the lattice reproduces in a sensible 
way the continuum 
results even if we use just first order (in $z$) results of the 
strong coupling perturbation theory. 

Using the value of $t$ given in Eq.(\ref{tequ}) one gets 
for the isotriplet mass Eq.(\ref{mtt2})
\begin{equation}
\frac{m_{T}}{e_{L}}=0.5143\quad .
\label{lb}
\end{equation}
By direct computation on an 8 sites chain one gets
\begin{equation}
\frac{m_{T}}{e_{L}}=1.3524\quad .
\label{ub}
\end{equation}
The discrepancy between Eq.(\ref{lb}) and Eq.(\ref{ub}) 
is mainly due to the approximation involved in the computation of 
$<g.s.|\vec{V}^{2}|g.s.>$. 
However, it is safe to believe that our lattice computation 
implies the existence of a massive isotriplet 
$1^{-+}$ with a mass between the lower bound (\ref{lb}) and 
the upper bound (\ref{ub}). This is in agreement 
with the results provided for the continuum theory in \cite{hara}. 
 
Using a similar procedure one may also compute the masses of the singlet $0^{++}$ and the triplet $1^{+-}$. 
From Eqs.(\ref{ms5},\ref{mt5}) one gets
\begin{eqnarray}
\frac{m_{S^{5}}}{e_{L}}&=&1.4290\quad .\\
\frac{m_{T^{5}}}{e_{L}}&=&1.3347\quad .
\end{eqnarray}
The triplet $|T^{5}>$ , being $G$-odd, is a scattering state of a $0^{--}$ 
singlet with a $1^{-+}$ triplet, which are the fundamental 
excitations of the system. 
The mass of this $1^{+-}$ triplet should be larger than the 
mass of the massive $1^{-+}$ triplet, which should be a 
scattering state of massless 
$1^{-+}$ triplets.

Putting $t=c=1$, $i.e.$ $e_{L}a=\frac{1}{z^{\frac{1}{4}}}$ 
the lattice mass spectrum gets closer to its continuum counterpart;
 for the isosinglet mass, one gets 
\begin{equation}
\frac{m_{S}}{e}=0.9938
\label{t1s}
\end{equation}
while for the isotriplet one gets 
\begin{equation}
\frac{m_{T}}{e_{L}}=0.4695\quad .
\label{mtt2}
\end{equation}
Eq.(\ref{t1s}) provides a value of the isosinglet mass lying 
$24\%$ above the exact answer. 
Again the triplet mass is reproduced with lesser accuracy due to 
the random phase approximation used in the computation of the 
pertinent correlator; 
a better answer is given however by a direct computation on the 
8 sites chain yielding the value $1.2346$ for ${m_{T}}/{e_{L}}$. 

\section{Concluding remarks}

In this paper we used the correspondence between the 2-flavor strongly 
coupled lattice Schwinger model and the antiferromagnetic Heisenberg 
Hamiltonian established in~\cite{noi2} to investigate the spectrum of 
the gauge model. Using the analysis
of the excitations of the finite size chains given in the appendix,
we showed the equality of the quantum numbers of the states of the Heisenberg 
model and the low-lying excitations of the 2-flavor Schwinger model.
We provided also the 
spectrum of the massive excitations of the gauge model; 
in order to extract numerical values for the masses, we explicitly computed  
the pertinent spin-spin correlators of the Heisenberg chain. 
Although the spectrum is determined  only up to the 
second order in the strong coupling expansion
the agreement with the continuum theory is satisfactory. 
 
The massless and the massive excitations of the gauge 
model are created from the 
spin chain ground state with two very different mechanisms: 
massless excitations involve only spin flipping 
while massive excitations are created by fermion transport 
besides spin flipping and do not belong to the spin chain spectrum.
As in the continuum theory, due to the Coleman 
theorem \cite{coleman}, the massless excitations are not Goldstone bosons,
but may be regarded as the gapless quantum excitations 
of the spin-$\frac{1}{2}$ antiferromagnetic 
Heisenberg chain \cite{haldane}. 

In computing the chiral condensate we showed that, also 
in the lattice theory, 
the expectation value of the umklapp operator $F$ is 
different from zero, while both 
$<\overline{\psi}\psi>$ and $<\overline{\psi}\sigma^{a}\psi>$ 
are zero to every order in the 
strong coupling expansion. 
This implies that both on the lattice and the continuum 
the $SU(2)$ flavor symmetry is preserved whereas the
$U_A(1)$ axial symmetry is broken. The umklapp operator $F$ is 
the order parameter for this symmetry, but being 
quadri-linear in the fermi fields, is invariant, in the continuum, under 
chiral rotation of $\pi/2$ and on the lattice under the corresponding 
discrete axial symmetry (\ref{chir}) (translation by one lattice site). 
This shows that the discrete axial 
symmetry is not broken in both cases.
Our lattice computation enhance this result since the ground state of the 
strongly coupled 2-flavor Schwinger model is translationally invariant.

The pattern of symmetry breaking of the continuum is 
exactly reproduced even if the Coleman theorem does not 
apply on the lattice and the anomalous symmetry breaking 
is impossible due to the Nielsen-Ninomiya~\cite{nini} theorem. 
At variance with the strongly coupled 1-flavor lattice 
Schwinger model, the anomaly 
is not realized in the lattice theory via the spontaneous breaking of 
a residual chiral symmetry~\cite{noi}, but, rather, 
by explicit breaking of the chiral symmetry due to staggered fermions.
The non-vanishing of $<F>$ may be regarded as
the only relic, in the strongly coupled 
lattice theory, of the anomaly of the 
continuum 2-flavor Schwinger model. 
It is due to the coupling induced by the 
gauge field, between the right and left-movers on the lattice.

It is quite straightforward to generalize our analysis to 
an $SU({\cal N})$-flavor group. 
For this, one should observe \cite{semen} that the results 
are very different depending on if ${\cal N}$ is even or odd. 
When ${\cal N}$ is odd, the ground state energy $-$ in the strong 
coupling limit $-$ is proportional to 
$e^2$ but,  
when ${\cal N}$ is even, the ground state energy is of order 1. 
This difference arises since, on the lattice, the charge density 
operator is odd under charge conjugation; 
therefore, the constant  $\frac{\cal N}{2}$ should be 
subtracted from the charge density operator \cite{semen}. 
As a consequence, when ${\cal N}$  is odd, the ground state 
supports electric fluxes while this becomes impossible when  
${\cal N}$  is even. 

When the fermion mass $m$ is different from zero, some 
further difference arises between ${\cal N}$ odd and ${\cal N}$ even. 
When ${\cal N}$ is odd, the mass term induces a translational
non invariant ground state, 
generating a spontaneous chiral symmetry breaking. 
When ${\cal N}$ is even, the ground state remains translationally
invariant in the strong coupling 
limit, $i.e.$ $e^2 \gg m$. 
In the weak coupling limit, $m\gg e^2$, the discrete chiral symmetry
is spontaneously broken 
for every ${\cal N}$. For ${\cal N}=2$, the soliton-antisoliton
excitations \cite{coleman} 
acquire a mass. 
\vskip 0.3truein  
{\large \bf Acknowledgements}

The authors are grateful to V. E. Korepin for making 
them aware of Ref.\cite{taka}, which revealed itself very valuable 
for the topic of section 5. 
We thank K. Harada for useful correspondence concerning the 
continuum spectrum of the 2-flavor Schwinger model. 
We thank also I. Affleck and L. D. Faddeev for valuable comments. 
This work has been supported by grants from the 
Natural Sciences and Engineering Research 
Council  of Canada, 
the Istituto Nazionale di Fisica Nucleare, and M.U.R.S.T. .   

\vskip 0.3truein 
\noindent
{\Large \bf Appendix A:}\\
{\Large \bf Exact diagonalization of finite size antiferromagnetic 
Heisenberg chains}
\vskip 0.3truein 
In the following we shall provide the exact diagonalization of the 
Heisenberg antiferromagnetic 
model for finite size chains of $N=4,6$ and $8$ sites and compare the results 
with the Bethe Ansatz solution provided in \cite{faddeev}. 
In order to make our arguments self-contained, we shall outline the steps 
involved in deriving the exact solution in the thermodynamic limit. 
We shall show that already very small finite size chains exhibit 
spectra that match very well with the thermodynamic limit solution. 
Furthermore the analysis of finite size chains is very useful since it allows 
the comparison between the quantum numbers of the Schwinger 
and the Heisenberg model excitations.
The interested reader may also look up references \cite{faddeev, lh, mattis}.

The one dimensional isotropic Heisenberg model describes a system of 
$N$ interacting spin-$\frac{1}{2}$ particles. 
The Hamiltonian of the model is
\begin{equation}
H_{J}=J\sum_{x=1}^{N}(\vec{S}_{x}\cdot \vec{S}_{x+1}-\frac{1}{4})\quad .
\label{heis}
\end{equation}
where $J>0$ ($J<0$ would describe a ferromagnet) and the spin operators 
have the following form
\begin{equation}
\vec{S}_{x}=1_{1}\otimes 1_{2}\otimes \ldots \otimes 
\frac{\vec{\sigma}_{x}}{2} 
\otimes \ldots \otimes 1_{N}\quad .
\end{equation}
They act nontrivially only on the Hilbert space of the $x^{th}$ site. 
Periodic boundary conditions are assumed. 

The Hamiltonian (\ref{heis}) is invariant under global rotations in the 
spin space, generated by
\begin{equation}
\vec{S}=\sum_{x=1}^{N}\vec{S}_{x}\quad .
\end{equation}
Due to the periodic boundary conditions, under translations generated by 
the operator $\hat{T}$, 
\begin{equation}
\hat{T}\quad : \vec{S}_{x} \longrightarrow \vec{S}_{x-1} 
\end{equation}
the Hamiltonian is invariant and $[\vec{S},\hat{T}]=0$.

The energy and the momentum of a given state 
with $M$ spins down can be written as~\cite{bethe}
\begin{eqnarray}
E_{M}&=&\sum_{\alpha=1}^{M}\epsilon_{\alpha}=-\frac{J}{2}
\sum_{\alpha=1}^{M}\frac{1}{\lambda_{\alpha}^{2}+\frac{1}{4}}\\
P_{M}&=&i\ln T=\sum_{\alpha=1}^{M}p_{\alpha}=i\sum_{\alpha=1}^{M}
\ln \frac{\lambda_{\alpha}-\frac{i}{2}}{\lambda_{\alpha}+\frac{i}{2}}\quad .
\end{eqnarray}
Energy and momentum are thus additive as if there were $M$ independent 
particles and the parameters $\lambda_{\alpha}$ must satisfy the  
Bethe Ansatz equations 
\begin{equation}
(\frac{\lambda_{\alpha}-\frac{i}{2}}{\lambda_{\alpha}+\frac{i}{2}})^{N}=-
\prod_{\beta=1}^{M}\frac{\lambda_{\alpha}-\lambda_{\beta}-i}
{ \lambda_{\alpha}-\lambda_{\beta}+i}\quad ,
\label{baeq}
\end{equation}
in order for $E_M$ and $P_M$ to be eigenvalues of the Hamiltonian
and momentum operator.
 See \cite{faddeev} for a derivation of Eqs.(\ref{baeq}).
 
\begin{figure}[htb]
\begin{center}
\setlength{\unitlength}{0.00041700in}%
\begingroup\makeatletter\ifx\SetFigFont\undefined
\def\x#1#2#3#4#5#6#7\relax{\def\x{#1#2#3#4#5#6}}%
\expandafter\x\fmtname xxxxxx\relax \def\y{splain}%
\ifx\x\y   
\gdef\SetFigFont#1#2#3{%
  \ifnum #1<17\tiny\else \ifnum #1<20\small\else
  \ifnum #1<24\normalsize\else \ifnum #1<29\large\else
  \ifnum #1<34\Large\else \ifnum #1<41\LARGE\else
     \huge\fi\fi\fi\fi\fi\fi
  \csname #3\endcsname}%
\else
\gdef\SetFigFont#1#2#3{\begingroup
  \count@#1\relax \ifnum 25<\count@\count@25\fi
  \def\x{\endgroup\@setsize\SetFigFont{#2pt}}%
  \expandafter\x
    \csname \romannumeral\the\count@ pt\expandafter\endcsname
    \csname @\romannumeral\the\count@ pt\endcsname
  \csname #3\endcsname}%
\fi
\fi\endgroup
\begin{picture}(8712,7514)(2701,-7718)
\thicklines
\put(6001,-5161){\circle{300}}
\put(7201,-1561){\circle{300}}
\put(7201,-6361){\circle{300}}
\put(7201,-3961){\circle{300}}
\put(8401,-2761){\circle{300}}
\put(8401,-361){\circle{300}}
\put(8401,-5161){\circle{300}}
\put(8401,-7561){\circle{300}}
\put(6001,-2761){\circle{300}}
\put(4801,-3961){\circle{300}}
\put(3601,-7561){\vector( 0, 1){7200}}
\put(10801,-4561){\makebox(0,0)[lb]{\smash{\SetFigFont{10}{12.0}{rm}Re$\lambda$}}}
\put(3601,-3961){\vector( 1, 0){7800}}
\multiput(6001,-2986)(0.00000,-7.98817){254}{\line( 0,-1){  3.994}}
\multiput(7201,-1786)(0.00000,-7.98817){254}{\line( 0,-1){  3.994}}
\multiput(7201,-4186)(0.00000,-7.98817){254}{\line( 0,-1){  3.994}}
\multiput(8401,-511)(0.00000,-8.00000){263}{\line( 0,-1){  4.000}}
\multiput(8401,-2911)(0.00000,-8.00000){263}{\line( 0,-1){  4.000}}
\multiput(8401,-5386)(0.00000,-7.98817){254}{\line( 0,-1){  3.994}}
\put(2701,-511){\makebox(0,0)[lb]{\smash{\SetFigFont{10}{12.0}{rm}Im$\lambda$}}}
\end{picture}
\end{center}
\caption{Strings for \protect $L=0,\frac{1}{2},1,\frac{3}{2}$}
\label{strings}
\end{figure}
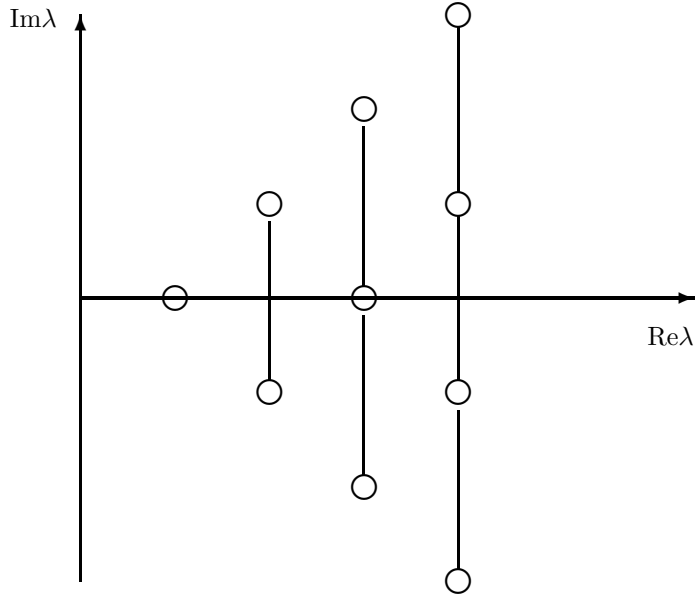

The solution of the antiferromagnetic Heisenberg chain is 
reduced to the solution of the system of the $M$ algebraic equations 
(\ref{baeq}). This, in general, is not an easy task.
It can be shown~\cite{faddeev}, however, that, in the thermodynamic limit 
$N\to\infty$, the complex parameters $\lambda$ have the form
\begin{equation}
\lambda_{l}=\lambda_{j,L}+il\quad,\quad l=-L,-L+1,\dots,L-1,L;
\label{complex}
\end{equation}
where $L$ is a non-negative integer or half-integer, $\lambda_{i,L}$
is the real part of the solution of (\ref{baeq}) and we shall 
define shortly the set of allowed values for the integer index $j$.
The $\lambda$'s that, for a given $\lambda_{j,L}$, are obtained varying $l$ 
between  $[-L,L]$ by integer steps, form a string of length $2L+1$,
see fig.(\ref{strings}). 
This arrangement of $\lambda$'s in the complex plane is called the 
``string hypothesis" \cite{faddeev}. 
In the following we shall verify 
that, even on finite size systems, the ``string hypothesis" 
is very well fulfilled. 

In a generic Bethe state with $M$ spins down, 
there are $M$ solutions to (\ref{baeq}), which can be grouped
according to the length of their strings.
Let us denote  by $\nu_{L}$ the number of strings of
length $2L+1$, $L=0,\frac{1}{2},\ldots$; strings of the same length
are obtained by changing the real parts, $\lambda_{j,L}$, of the $\lambda$'s 
in (\ref{complex}); as a consequence $\quad j=1,\ldots,\nu_{L}$. 
If one denotes the total number of strings by $q$ one has
\begin{equation}
q=\sum_{L}\nu _{L}\quad,\quad M=\sum_{L} (2L+1)\nu _{L}\quad .
\label{constraint}
\end{equation}

The set of integers $(M,q,\{\nu_{L}\})$ constrained 
by (\ref{constraint}), characterizes Bethe 
states up to the fixing of the $q$ numbers $\lambda_{j,L}$; this set 
is called the ``configuration''. 
Varying $M$, $q$ and $\nu_L$, one is able to construct all the
$2^N$ eigenstates of an Heisenberg antiferromagnetic chain of $N$ 
sites~\cite{faddeev}.
The energy and momentum of the Bethe's state, corresponding to a 
given configuration $-$ within exponential accuracy as
$N\rightarrow \infty$ $-$ consist of $q$ summands representing 
the energy and momentum of separate strings. 
For the parameters $\lambda_{j,L}$ of the given configuration, 
taking the logarithm of (\ref{baeq}) 
the following system of equations is obtained in the thermodynamic limit
\begin{equation}
2N\arctan \frac{\lambda_{j,L_{1}}}{L_{1}+\frac{1}{2}}=2\pi 
Q_{j,L_{1}}+\sum_{L_{2}}
\sum_{k=1}^{\nu_{L_{2}}}\Phi_{L_1 L_2}(\lambda_{j,L_{1}}-
\lambda_{k,L_{2}})\quad, 
\label{tbaeq}
\end{equation}
where
\begin{equation}  
\Phi_{L_1 L_2}(\lambda)=2\sum_{L=|L_1 -L_2|\neq 0}^{L_1 +L_2}
(\arctan \frac{\lambda}{L}+\arctan \frac{\lambda}{L+1})\quad .
\end{equation}

Integer and half integer numbers $Q_{j,L}$ parametrize the 
branches of the arcotangents and, consequently, the 
possible solutions of the system of Eqs.(\ref{tbaeq}). In ref.\cite{faddeev}
it was shown that the $Q_{j,L}$  are limited as
\begin{equation}
-Q_{L}^{max}\leq Q_{1,L}<Q_{2,L}<\ldots<Q_{\nu_{L},L}\leq Q_{L}^{max}\quad 
\label{vacan}
\end{equation}
with $Q_{L}^{max}$ given by
\begin{equation}
Q_{L}^{max}=\frac{N}{2}-\sum_{L'}J(L,L')\nu_{L'}-\frac{1}{2}
\label{qmax}
\end{equation}
and
\begin{equation}
 J(L_1 ,L_2)=\left\{\begin{array}{ll}
 2{\rm min}(L_1 ,L_2)+1 &\mbox{if $L_1 \neq L_2$}\\ 
 2L_1+\frac{1}{2} &\mbox{if $L_1 =L_2$}\quad .
\end{array}
\right.
\label{jll}
\end{equation}
The admissible values for the numbers $Q_{j,L}$ are called 
the ``vacancies'' and their number for every $L$ is denoted by $P_{L}$
\begin{equation}
P_{L}=2Q_{L}^{max}+1\quad .
\label{pelle}
\end{equation}

The main hypothesis formulated in~\cite{faddeev} is that to every 
admissible collection of $Q_{j,L}$ there corresponds a unique 
solution of the system of equations (\ref{tbaeq}). 
The solution always provides, in a multiplet, the state with the highest
value of the third spin component $S^3$. 

Let us now consider some simple example.
The simplest configuration has only strings of length 1,
$i.e.$ all the $\lambda$'s are real. 
The singlet associated to this configuration
\begin{equation}
M=q=\nu_0 =\frac{N}{2}\quad,\quad \nu_{L}=0\quad,\quad L>0\quad ,
\end{equation}
is the ground state.
The vacancies of the strings of length 1, $i.e.$ the 
admissible values of $Q_{j,0}$, due to 
eqs.(\ref{vacan},\ref{qmax},\ref{jll}), belong to the segment 
\begin{equation}
-\frac{N}{4}+\frac{1}{2}\leq Q_{j,0}\leq \frac{N}{4}-\frac{1}{2}\quad .
\label{igs}
\end{equation}
Therefore they are $N/2$.
All these vacancies must then be used to find the $N/2$ strings of length 1.
As a consequence this state is uniquely specified and no degeneracy 
is possible.

Next we consider the configuration that provides a singlet
with 1 string of length 2 and all the others of length 1:
\begin{equation}
M=\frac{N}{2}\quad,\quad q=\frac{N}{2}-1\quad,\quad \nu_{0}=
\frac{N}{2}-2 \quad,\quad \nu_{\frac{1}{2}}=1\quad,\quad \nu_{L}
=0\quad,\quad L>\frac{1}{2}\ .
\label{c1}
\end{equation}
For the strings of length 1 the 
number of vacancies is again ${N}/{2}$; 
for the string of length 2 there is one vacancy and the only 
admissible $Q_{j,1}$ equals 0. 
Thus, since the number of strings of length 1 is $\nu_{0}=
\frac{N}{2}-2$, there are two vacancies for which Eqs.(\ref{tbaeq}) 
have no solution; they are called ``holes'' and are denoted $Q_{1}^{(h)}$ 
and $Q_{2}^{(h)}$. 
This configuration is determined by two parameters: 
the positions of two ``holes" which vary independently in the interval 
(\ref{igs}). 

There is another state with only 2 holes:
the triplet corresponding to the configuration
\begin{equation}
M=q=\nu_{0}=\frac{N}{2}-1\quad,\quad \nu_{L}=0\quad,\quad L>0
\label{c2}
\end{equation}
The number of vacancies for the strings of length 1 equals $\frac{N}{2}+1$, 
while $\nu_{0}=\frac{N}{2}-1$.

The excitations determined by the configurations 
(\ref{c1},\ref{c2}) belong to the configuration class called in~\cite{faddeev}
$\cal{M}_{AF}$. The class $\cal{M}_{AF}$ is characterized as follows: 
the number of strings of length 1 in each configuration 
belonging to this class
differs by a finite quantity from ${N}/{2}$, $\nu_0=\frac{N}{2}-k_0$ 
where $k_0$
is a positive finite constant, so that 
the number of strings of length greater than 1 is finite. From (\ref{pelle}) 
we then have
\begin{eqnarray}
P_0&=&\frac{N}{2}+k_0-2\sum_{L>0}\nu_L\label{p0}\\
P_L&=&2k_0-2\sum_{L'>0}J(L,L')\nu_{L'}\ ,\quad L>0
\end{eqnarray}
so that
\begin{equation}
P_0\ge \frac{N}{2}\ ,\quad P_L<2 k_0\ ,\quad L>0\ .
\end{equation}
From (\ref{p0}) follows that the number of holes for the strings of length 1 
is always even and equals 2 only for the singlet and the triplet 
excitations discussed above.
One can imagine the class $\cal{M}_{AF}$ as a
``sea" of strings of length 1 with a finite number of 
strings of length greater than 1 immersed into it. 
It was proven in~\cite{faddeev} that, in the 
thermodynamic limit, the states belonging to 
$\cal{M}_{AF}$ have finite energy and momentum with 
respect to the antiferromagnetic vacuum, whereas 
each of the states which 
corresponds to a configuration not included in the class $\cal{M}_{AF}$ has 
an infinite energy relative to the antiferromagnetic vacuum.

Let us now sketch the computation of the thermodynamic limit 
ground state energy. Eqs.(\ref{tbaeq}) for the ground state have the form
\begin{equation}
\arctan 2\lambda_{j}=\frac{\pi Q_{j}}{N}+\frac{1}{N}
\sum_{k=1}^{{N}/{2}}\arctan (\lambda_{j} -\lambda_{k})\quad .
\label{gstbae}
\end{equation}
Taking the thermodynamic limit $N\rightarrow \infty$, one has 
\begin{equation}
\frac{Q_{j}}{N}\rightarrow x\quad,\quad -\frac{1}{4}\leq x\leq 
\frac{1}{4}\quad,\quad \lambda_{j}\rightarrow \lambda(x)\quad ,
\end{equation}
and Eqs.(\ref{gstbae}) can be rewritten in the form
\begin{equation}
\arctan 2\lambda(x)=\pi x+\int_{-\frac{1}{4}}^{\frac{1}{4}}
\arctan(\lambda(x)-\lambda(y)) dy\quad .
\label{2gsbae}
\end{equation}
Upon introducing the density of the numbers $\lambda(x)$ 
in the interval $d\lambda$ 
\begin{equation}
\rho(\lambda)=\frac{1}{\frac{d\lambda(x)}{dx}|_{x=x(\lambda)}}
\label{den}
\end{equation}
and differentiating Eqs.(\ref{2gsbae}), one gets
\begin{equation}
\rho(\lambda)=\frac{1}{2\pi}\int_{-\infty}^{\infty}
\frac{e^{-\frac{1}{2}|\xi|}}{1+e^{-|\xi|}}e^{-i\lambda |\xi|}
d\xi =\frac{1}{2\cosh\pi\lambda}\quad .
\label{den1}
\end{equation}
The density $\rho(\lambda)$ introduced in this way is normalized to $1/2$.
It is now easy to compute the energy and the momentum of the ground state 
\begin{equation}
E_{g.s.}=\sum_{\alpha =1}^{\frac{N}{2}}\epsilon_{\alpha}
=N\int_{-\infty}^{\infty}\epsilon(\lambda)\rho(\lambda)d\lambda=
-\frac{J N}{4}\int_{-\infty}^{\infty}d\lambda
\frac{1}{\left(\lambda^2+\frac{1}{4}
\right)\cosh\pi\lambda}=-JN\ln 2
\label{gse}
\end{equation}
\begin{equation}
P_{g.s.}=\sum_{\alpha=1}^{\frac{N}{2}}p_{\alpha}=N
\int_{-\infty}^{\infty}p(\lambda)
\rho(\lambda)d\lambda=-\frac{N}{2}
\int_{-\infty}^{\infty}d\lambda\frac{\pi}{\cosh\pi\lambda}=
\frac{N}{2}\pi\quad ({\rm mod}\quad 2\pi)\quad .
\label{gsm}
\end{equation}
According to Eq.(\ref{gsm}), $P_{g.s.}=0\ ({\rm mod} 2\pi)$ 
for $\frac{N}{2}$ even, and 
$P_{g.s.}=\pi\ ({\rm mod}\  2\pi)$ for $\frac{N}{2}$ odd. 
The ground state, as expected, is a singlet, in fact the spin $S$ 
is given by
\begin{equation}
S=\frac{N}{2}-\sum_{\alpha=1}^{N/2}1=\frac{N}{2}-
N\int_{-\infty}^{\infty}\rho(\lambda) d\lambda=0\quad .
\end{equation}

Let us analyze the triplet described by Eq.(\ref{c2}); 
Eqs.(\ref{tbaeq}) take the form
\begin{equation}
\arctan 2\lambda_{j}=\frac{\pi Q_{j}}{N}+\frac{1}{N}
\sum_{k=1}^{\frac{N}{2}-1}\arctan (\lambda_{j} -\lambda_{k})
\label{trbae}
\end{equation}
where now the numbers $Q_{j}$ lie in the segment 
$[-\frac{N}{4},\frac{N}{4}]$ and have two holes, $Q_1 ^{(h)}$ and 
$Q_2 ^{(h)}$ with $Q_1 ^{(h)} <Q_2 ^{(h)}$. Taking the 
thermodynamic limit one gets
\begin{equation}
\frac{Q_1 ^{(h)}}{N}\rightarrow x_1\quad,\quad 
\frac{Q_2 ^{(h)}}{N}\rightarrow x_2\quad,\quad \frac{Q_j}{N}
\rightarrow x+\frac{1}{N}(
\theta(x-x_1)+\theta(x-x_2))
\end{equation}
where $\theta (x)$ is the Heaviside function. Eqs.(\ref{trbae}) become 
\begin{equation}
\arctan 2\lambda(x)=\pi x+\frac{\pi}{N}(\theta(x-x_1)+\theta(x-x_2)) +
 \int_{-\frac{1}{4}}^{\frac{1}{4}}\arctan(\lambda(x)-\lambda(y)) dy\quad .
\label{2trbae}
\end{equation}
Eq.(\ref{2trbae}) gives, for this triplet, the density of 
$\lambda$, $\rho(\lambda)=\frac{d \lambda}{dx}$ 
\begin{equation}
\rho_{t}(\lambda)=\rho(\lambda)+\frac{1}{N}(\sigma (\lambda -
\lambda_1)-\sigma (\lambda -\lambda_2))
\end{equation}
where $\rho (\lambda)$ is given in Eq.(\ref{den1}) and  
\begin{equation}
\sigma(\lambda )=-\frac{1}{2\pi }\int_{-\infty}^{\infty}\frac{1}{1+e^{-|\xi|}}
e^{-i\lambda \xi} d\xi \quad .
\label{densig}
\end{equation}
$\lambda_1$ and $\lambda_2$ are the parameters of the holes, 
$\lambda_i=\lambda(x_{i})$, $i=1,2$.
The energy and the momentum of this state measured from the 
ground state are now easily computed
\begin{equation}
\epsilon_{T}(\lambda_1,\lambda_2)=
N\int_{-\infty}^{\infty}\epsilon(\lambda) 
(\rho_{t}(\lambda)-\rho(\lambda))d\lambda =
\epsilon(\lambda_1)+\epsilon(\lambda_2)
\label{tre}
\end{equation}
\begin{equation}
p_{T}(\lambda_1 ,\lambda_2)=N\int_{-\infty}^{\infty}p(\lambda) 
(\rho_{t}(\lambda)-\rho(\lambda))d\lambda=
p(\lambda_1)+p(\lambda_2)\quad (mod\quad 2\pi)
\end{equation}
where
\begin{equation}
\epsilon(\lambda)=\int_{-\infty}^{\infty}\epsilon(\mu)
\sigma(\lambda -\mu)d\mu=J\frac{\pi}{2\cosh \pi\lambda}
\label{et}
\end{equation}
\begin{equation}
p(\lambda)=\int_{-\infty}^{\infty}p(\mu)\sigma(\lambda -\mu)
d\mu=\arctan \sinh \pi \lambda -\frac{\pi}{2},\quad -\pi
\leq p(\lambda) \leq 0\quad .
\label{pt}
\end{equation}
From Eqs.(\ref{et},\ref{pt}) one gets
\begin{equation}
\epsilon=-\frac{J\pi}{2}\sin p\quad .
\label{disrel}
\end{equation}
The momentum $p_T(\lambda_1,\lambda_2)$ varies over 
the interval $[0,2\pi)$, when $\lambda_1$ and $\lambda_2$ run 
independently over the whole real axis.
The spin of this state can be computed by the formula
\begin{equation}
S=-\int_{-\infty}^{\infty}(\sigma(\lambda -\lambda_1)+
\sigma(\lambda -\lambda_2))d\lambda=1\quad .
\end{equation}

Let us finally analize the singlet excitation 
characterized by the configuration (\ref{c1}). 
Denoting by $\lambda_S$ 
the only number among the $\lambda_{j,{1}/{2}}$ 
which characterizes the string of length 2 and by $\lambda_{j}$ 
the numbers $\lambda_{j,0}$ 
for the strings of length 1, Eqs.(\ref{tbaeq}) read
\begin{eqnarray}
\arctan 2\lambda_{j}&=&\frac{\pi Q_j}{N}+\frac{1}{N}
\Phi(\lambda_j -\lambda_S)+\frac{1}{N}
\sum_{k=1}^{\frac{N}{2}-2}\arctan (\lambda_j -\lambda_k)\\
\arctan \lambda_S&=&\frac{1}{N}
\sum_{j=1}^{\frac{N}{2}-2}\Phi(\lambda_S -\lambda_j)
\end{eqnarray}
with
\begin{equation}
\Phi(\lambda)=\arctan 2\lambda+\arctan \frac{2}{3}\lambda\quad .
\end{equation}
The $\frac{N}{2}-2$ numbers $Q_{j}$ vary in the segment 
$[-\frac{N}{4}+\frac{1}{2},\frac{N}{4}-\frac{1}{2}]$; 
among them there are the two holes $Q_{1}^{(h)}$ and 
$Q_{2}^{(h)}$. Taking the thermodynamic limit one finds 
the density of $\lambda$'s for the singlet
\begin{equation}
\rho(\lambda_{S})=\rho(\lambda)+\frac{1}{N}
(\sigma(\lambda -\lambda_1)+\sigma(\lambda -\lambda_2)+
\omega(\lambda -\lambda_S))
\label{dens}
\end{equation}
where $\rho$ and $\sigma$ were given in Eqs.(\ref{den1}, 
\ref{densig}) and where 
\begin{equation}
\omega(\lambda)=-\frac{1}{2\pi}\int_{-\infty}^{\infty}
e^{-\frac{1}{2}|\xi|-i\lambda \xi} d\xi=
-\frac{2}{\pi(1+4\lambda^2)}\quad .
\end{equation}
In \cite{faddeev} it was demonstrated that the 
string parameter $\lambda_S$ is uniquely determined by 
the $\lambda$'s parametrizing the two holes
\begin{equation}
\lambda_S=\frac{\lambda_{1}^{(h)}+\lambda_{2}^{(h)}}{2}\quad .
\end{equation}
In \cite{faddeev} it was also proved the remarkable fact 
that the string of length 2 does not contribute to the 
energy and momentum of the excitation, 
so that the singlet and the triplet have the same dispersion relations
\begin{eqnarray}
\epsilon_S(\lambda_1,\lambda_2)&=&\epsilon_{T}(\lambda_1,\lambda_2)
=\epsilon(\lambda_1)+\epsilon(\lambda_2) \\
p_S(\lambda_1 ,\lambda_2)&=&p_T(\lambda_1 ,\lambda_2)=
p(\lambda_1)+p(\lambda_2)\quad ({\rm mod}\quad 2\pi)\quad .
\end{eqnarray}
The spin of this excitation is, of course, zero
\begin{equation}
S=-2-\int_{-\infty}^{\infty}(2\sigma(\lambda)+\omega(\lambda)) d\lambda=0
\end{equation}
The only difference between the state 
whose configuration is given in Eq.(\ref{c2}) and the state of
Eq.(\ref{c1}) is the spin. 

To summarize,
the finite energy 
excitations of the antiferromagnetic Heisenberg chain are only 
those belonging to the class ${\cal M_{AF}}$ and are described by
scattering states of an even number of quasiparticles or kinks. 
The momentum $p$ of these kinks runs over half the Brillouin 
zone $-\pi\le p\le 0$, the 
dispersion relation is $\epsilon(p)=\frac{J\pi}{2}\sin p$, Eq.(\ref{disrel}),
and the spin of a kink is $1/2$. The singlet and the triplet excitations
described above are the only states composed of two kinks, the spins of the 
kinks being parallel for the triplet and antiparallel for the singlet.
For vanishing total momentum all the states belonging to ${\cal M_{AF}}$
have the same energy of the ground state so that they are gapless excitations.
Since the eigenstates of $H_{J}$ always contain an even number of kinks,
the dispersion relation is determined by a a set of two-parameters: the
momenta of the even number of kinks whose scattering provides the excitation. 
There are no bound states of kinks.

Let us now turn to the computation of the spectrum of 
finite size quantum antiferromagnetic chains 
by exact diagonalization. We shall see that already 
for very small chains, the spectrum is well described  
by the Bethe ansatz solution in the thermodynamic limit. 
Furthermore, an intuitive picture 
of the ground state and of the lowest lying excitations
of the strongly coupled 2-flavor 
lattice Schwinger model emerges. 
 
The states of an antiferromagnetic chain are 
classified according to the quantum numbers of spin, 
third spin component, energy and momentum $|S, S^3,E,p>$. 
For a 4 site chain the 
momenta allowed for the states are: $0,
\frac{\pi}{2},\frac{3\pi}{2}\ {\rm mod}\  2\pi $. The ground state is
\begin{equation}
|g.s.>=|0,0,-3J,0>=\frac{1}{\sqrt{12}}(2|\uparrow
\downarrow\uparrow\downarrow>+2|\downarrow\uparrow\downarrow\uparrow>-
|\uparrow\uparrow\downarrow \downarrow> -|\uparrow
\downarrow \downarrow\uparrow>-|\downarrow \downarrow\uparrow\uparrow>-
|\downarrow\uparrow\uparrow\downarrow>)\quad .
\label{gs4}
\end{equation}
This state is $P$-parity even.
In fact, by the definition of $P$-parity given in Eq.(\ref{par}),
the $P$-parity inverted state (\ref{gs4}) is
obtained by reverting the order of the spins in each vector $|\dots>$
appearing in (\ref{gs4}), e.g. $|\downarrow\downarrow\uparrow\uparrow>
\buildrel P\over\longrightarrow|\uparrow\uparrow\downarrow \downarrow>$.

The $\lambda$'s associated to the ground state (solution of the Bethe 
ansatz equations (\ref{baeq})) are 
$\lambda_1=-\frac{1}{2\sqrt{3}}$ and $\lambda_2=\frac{1}{2\sqrt{3}}$. 
There is also an excited singlet 
\begin{equation}
|0,0,-J,\pi>=\frac{1}{\sqrt{4}}(|\downarrow \downarrow
\uparrow\uparrow>-|\downarrow\uparrow\uparrow\downarrow>
-|\uparrow\downarrow \downarrow\uparrow>+ |\uparrow
\uparrow\downarrow \downarrow>)\quad .
\label{sex1}
\end{equation}
It is $P$-even, so that it is a $S^{P}=0^{+}$ excitation, with
the same quantum numbers (the isospin is replaced by the spin)
of the lowest lying singlet excitation of the
strongly coupled Schwinger model discussed by Coleman \cite{coleman}. 
The state (\ref{sex1}) also coincides with the excited 
singlet described by the configuration (\ref{c1}). It has only two 
complex $\lambda$'s which arrange 
themselves in a string approximately of length 2, 
$\lambda_1=-\lambda_2=i\sqrt{\frac{\sqrt{481}-17}{8}}$ and 
there are two holes with $Q_{1}^{(h)}=-\frac{1}{2}$ and 
$Q_{2}^{(h)}=\frac{1}{2}$. 

There are also three excited triplets, whose highest weight states are 
\begin{eqnarray}
|1,1,-J,\frac{\pi}{2}>&=&\frac{1}{\sqrt{4}}
(|\downarrow\uparrow\uparrow\uparrow>+i|\uparrow\downarrow\uparrow\uparrow>-
|\uparrow\uparrow\downarrow\uparrow>-i|\uparrow\uparrow\uparrow\downarrow>)\\
|1,1,-2J,\pi>&=&\frac{1}{\sqrt{4}}(|\downarrow
\uparrow\uparrow\uparrow>-|\uparrow\downarrow\uparrow\uparrow>+
|\uparrow\uparrow\downarrow\uparrow>-|\uparrow\uparrow\uparrow\downarrow>)
\label{tr63}\\
|1,1,-J,\frac{3\pi}{2}>&=&\frac{1}{\sqrt{4}}(|\downarrow
\uparrow\uparrow\uparrow>-i|\uparrow\downarrow\uparrow\uparrow>-
|\uparrow\uparrow\downarrow\uparrow>+i|\uparrow\uparrow\uparrow\downarrow>)\quad .
\label{6tri}
\end{eqnarray}
Among these, only the non-degenerate state with the lowest energy has a
well defined $P$-parity (\ref{tr63}). It is a $S^P=1^{-}$
like the lowest lying triplet of the 2-flavor strongly coupled Schwinger
model. The degenerate states can be always combined
to form a $P$-odd state. 

We thus see that within the states in a given 
configuration there is always a representative state with well defined parity,
the others are degenerate and can be used to construct state of well defined 
energy and parity. Moreover the parity of the representative states
(with respect to the parity of the ground state) is
the same of the one of the lowest-lying
Schwinger model excitations in strong coupling.

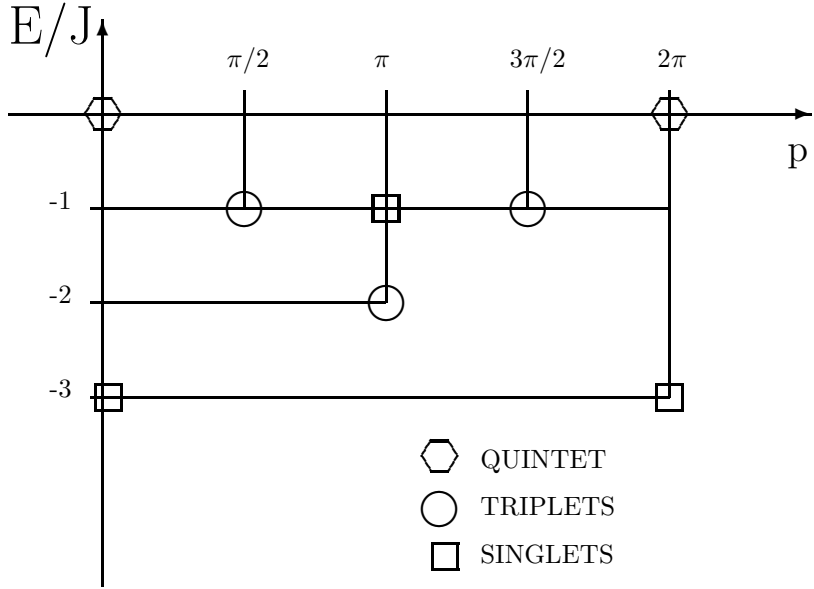
\begin{figure}[htb]
\begin{center}
\setlength{\unitlength}{0.00041700in}%
\begingroup\makeatletter\ifx\SetFigFont\undefined
\def\x#1#2#3#4#5#6#7\relax{\def\x{#1#2#3#4#5#6}}%
\expandafter\x\fmtname xxxxxx\relax \def\y{splain}%
\ifx\x\y   
\gdef\SetFigFont#1#2#3{%
  \ifnum #1<17\tiny\else \ifnum #1<20\small\else
  \ifnum #1<24\normalsize\else \ifnum #1<29\large\else
  \ifnum #1<34\Large\else \ifnum #1<41\LARGE\else
     \huge\fi\fi\fi\fi\fi\fi
  \csname #3\endcsname}%
\else
\gdef\SetFigFont#1#2#3{\begingroup
  \count@#1\relax \ifnum 25<\count@\count@25\fi
  \def\x{\endgroup\@setsize\SetFigFont{#2pt}}%
  \expandafter\x
    \csname \romannumeral\the\count@ pt\expandafter\endcsname
    \csname @\romannumeral\the\count@ pt\endcsname
  \csname #3\endcsname}%
\fi
\fi\endgroup
\begin{picture}(10224,7224)(1189,-6973)
\thicklines
\put(6001,-3361){\circle{450}}
\put(7801,-2161){\circle{450}}
\put(6676,-5986){\circle{450}}
\put(2401,-6961){\vector( 0, 1){7200}}
\put(1201,-961){\vector( 1, 0){10200}}
\put(2401,-2161){\line( 1, 0){7200}}
\put(2401,-2161){\line( 1, 0){7200}}
\put(2401,-3361){\line( 1, 0){3600}}
\put(2401,-4561){\line( 1, 0){ 75}}
\put(2251,-2161){\line( 1, 0){150}}
\put(2251,-3361){\line( 1, 0){150}}
\put(2251,-4561){\line( 1, 0){150}}
\put(4201,-661){\line( 0,-1){1500}}
\put(6001,-661){\line( 0,-1){2700}}
\put(7801,-661){\line( 0,-1){1500}}
\put(9601,-661){\line( 0,-1){300}}
\multiput(2626,-961)(-8.91855,14.86426){14}{\makebox(13.3333,20.0000){\SetFigFont{7}{8.4}{rm}.}}
\put(2513,-766){\line(-1, 0){224}}
\multiput(2289,-766)(-8.91855,-14.86426){14}{\makebox(13.3333,20.0000){\SetFigFont{7}{8.4}{rm}.}}
\multiput(2176,-961)(8.91855,-14.86426){14}{\makebox(13.3333,20.0000){\SetFigFont{7}{8.4}{rm}.}}
\put(2289,-1156){\line( 1, 0){224}}
\multiput(2513,-1156)(8.91855,14.86426){14}{\makebox(13.3333,20.0000){\SetFigFont{7}{8.4}{rm}.}}
\multiput(9826,-961)(-8.91855,14.86426){14}{\makebox(13.3333,20.0000){\SetFigFont{7}{8.4}{rm}.}}
\put(9713,-766){\line(-1, 0){224}}
\multiput(9489,-766)(-8.91855,-14.86426){14}{\makebox(13.3333,20.0000){\SetFigFont{7}{8.4}{rm}.}}
\multiput(9376,-961)(8.91855,-14.86426){14}{\makebox(13.3333,20.0000){\SetFigFont{7}{8.4}{rm}.}}
\put(9489,-1156){\line( 1, 0){224}}
\multiput(9713,-1156)(8.91855,14.86426){14}{\makebox(13.3333,20.0000){\SetFigFont{7}{8.4}{rm}.}}
\put(4201,-2161){\circle{450}}
\put(5851,-2311){\framebox(300,300){}}
\put(1726,-2161){\makebox(0,0)[lb]{\smash{\SetFigFont{10}{12.0}{rm}-1}}}
\put(2326,-4711){\framebox(300,300){}}
\multiput(6901,-5311)(-8.91855,14.86426){14}{\makebox(13.3333,20.0000){\SetFigFont{7}{8.4}{rm}.}}
\put(6788,-5116){\line(-1, 0){224}}
\multiput(6564,-5116)(-8.91855,-14.86426){14}{\makebox(13.3333,20.0000){\SetFigFont{7}{8.4}{rm}.}}
\multiput(6451,-5311)(8.91855,-14.86426){14}{\makebox(13.3333,20.0000){\SetFigFont{7}{8.4}{rm}.}}
\put(6564,-5506){\line( 1, 0){224}}
\multiput(6788,-5506)(8.91855,14.86426){14}{\makebox(13.3333,20.0000){\SetFigFont{7}{8.4}{rm}.}}
\put(6601,-6736){\framebox(300,300){}}
\put(2401,-2161){\line( 1, 0){5400}}
\put(2401,-4561){\line( 1, 0){7200}}
\put(9601,-961){\line( 0,-1){3600}}
\put(9451,-4711){\framebox(300,300){}}
\put(11101,-1561){\makebox(0,0)[lb]{\smash{\SetFigFont{14}{16.8}{rm}p}}}
\put(1726,-3361){\makebox(0,0)[lb]{\smash{\SetFigFont{10}{12.0}{rm}-2}}}
\put(1726,-4561){\makebox(0,0)[lb]{\smash{\SetFigFont{10}{12.0}{rm}-3}}}
\put(7201,-5461){\makebox(0,0)[lb]{\smash{\SetFigFont{10}{12.0}{rm}QUINTET}}}
\put(7201,-6061){\makebox(0,0)[lb]{\smash{\SetFigFont{10}{12.0}{rm}TRIPLETS}}}
\put(7201,-6661){\makebox(0,0)[lb]{\smash{\SetFigFont{10}{12.0}{rm}SINGLETS}}}
\put(3976,-361){\makebox(0,0)[lb]{\smash{\SetFigFont{10}{12.0}{rm}$\pi /2$}}}
\put(5851,-361){\makebox(0,0)[lb]{\smash{\SetFigFont{10}{12.0}{rm}$\pi$}}}
\put(7576,-361){\makebox(0,0)[lb]{\smash{\SetFigFont{10}{12.0}{rm}$3\pi /2$}}}
\put(9451,-361){\makebox(0,0)[lb]{\smash{\SetFigFont{10}{12.0}{rm}$2\pi$}}}
\put(1201,-61){\makebox(0,0)[lb]{\smash{\SetFigFont{20}{24.0}{rm}E/J}}}
\end{picture}
\end{center}
\caption{Four sites chain spectrum}
\label{4spectrum}
\end{figure}

All the triplets in (\ref{6tri}) have one real $\lambda$ and two holes; 
they can be associated with the family of triplets (\ref{c2}).
 In table (\ref{tiqn}) we summarize the triplet $\lambda$'s and $Q^{(h)}$'s.
\begin{table}[htbp]
\begin{center}
\caption{Triplet internal quantum numbers }\label{tiqn}
\vspace{.1in}
\begin{tabular}{|lccc|}
\hline
TRIPLET&$\lambda$&$Q_{1}^{(h)}$&$Q_{2}^{(h)}$\rule{0in}{4ex}\\[2ex] \hline
$|1,1,-J,\frac{\pi}{2}>$& $\frac{1}{2} $&$-1$&$0$\rule{0in}{4ex}\\[2ex] \hline
$|1,1,-2J,\pi>$&$0$&$-1$&$1$\rule{0in}{4ex}\\[2ex] \hline
$|1,1,-J,\frac{3\pi}{2}>$&$-\frac{1}{2}$&$0$&$1$\rule{0in}{4ex}\\[2ex] \hline
\end{tabular}
\end{center}
\end{table}
The spectrum exhibits also a quintet, whose highest weight state is 
\begin{equation}
|2,2,0,0>=|\uparrow\uparrow\uparrow\uparrow>
\end{equation}
We report in fig.(\ref{4spectrum}) the spectrum of the 4 sites chain. 

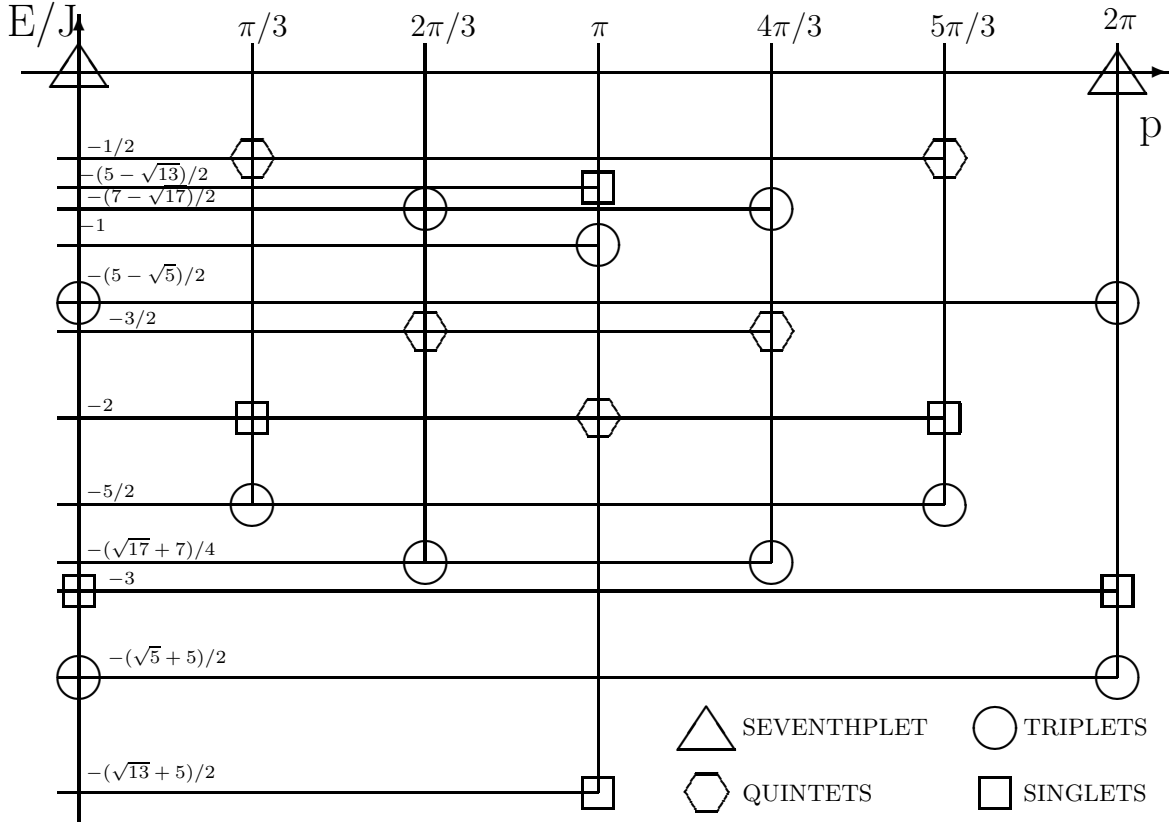
\begin{figure}[htb]
\begin{center}
\setlength{\unitlength}{0.00050000in}%
\begingroup\makeatletter\ifx\SetFigFont\undefined
\def\x#1#2#3#4#5#6#7\relax{\def\x{#1#2#3#4#5#6}}%
\expandafter\x\fmtname xxxxxx\relax \def\y{splain}%
\ifx\x\y   
\gdef\SetFigFont#1#2#3{%
  \ifnum #1<17\tiny\else \ifnum #1<20\small\else
  \ifnum #1<24\normalsize\else \ifnum #1<29\large\else
  \ifnum #1<34\Large\else \ifnum #1<41\LARGE\else
     \huge\fi\fi\fi\fi\fi\fi
  \csname #3\endcsname}%
\else
\gdef\SetFigFont#1#2#3{\begingroup
  \count@#1\relax \ifnum 25<\count@\count@25\fi
  \def\x{\endgroup\@setsize\SetFigFont{#2pt}}%
  \expandafter\x
    \csname \romannumeral\the\count@ pt\expandafter\endcsname
    \csname @\romannumeral\the\count@ pt\endcsname
  \csname #3\endcsname}%
\fi
\fi\endgroup
\begin{picture}(12087,8499)(451,-8173)
\thicklines
\put(12001,-6661){\circle{450}}
\put(4801,-5461){\circle{450}}
\put(8401,-5461){\circle{450}}
\put(3001,-4861){\circle{450}}
\put(10201,-4861){\circle{450}}
\put(1201,-2761){\circle{450}}
\put(12001,-2761){\circle{450}}
\put(6601,-2161){\circle{450}}
\put(4801,-1786){\circle{450}}
\put(8401,-1786){\circle{450}}
\put(10726,-7186){\circle{450}}
\put(1201,-8161){\vector( 0, 1){8400}}
\put(3001,-61){\line( 0,-1){300}}
\put(4801,-61){\line( 0,-1){300}}
\put(6601,-61){\line( 0,-1){300}}
\put(8401,-61){\line( 0,-1){300}}
\put(10201,-61){\line( 0,-1){300}}
\put(12001,-61){\line( 0,-1){300}}
\put(976,-2161){\line( 1, 0){5625}}
\put(6601,-361){\line( 0,-1){1800}}
\put(976,-1261){\line( 1, 0){9225}}
\put(3001,-361){\line( 0,-1){900}}
\put(10201,-361){\line( 0,-1){900}}
\put(976,-1561){\line( 1, 0){5625}}
\put(4801,-361){\line( 0,-1){1425}}
\put(8401,-361){\line( 0,-1){1425}}
\put(976,-3061){\line( 1, 0){7425}}
\put(976,-2761){\line( 1, 0){225}}
\put(4801,-1786){\line( 0,-1){1275}}
\put(8401,-1786){\line( 0,-1){1275}}
\put(3001,-1261){\line( 0,-1){2700}}
\put(6601,-2161){\line( 0,-1){1800}}
\put(10201,-1261){\line( 0,-1){2700}}
\put(976,-3961){\line( 1, 0){9225}}
\put(976,-4861){\line( 1, 0){9225}}
\put(3001,-3961){\line( 0,-1){900}}
\put(10201,-3961){\line( 0,-1){900}}
\put(976,-5761){\line( 1, 0){225}}
\put(976,-5461){\line( 1, 0){7425}}
\put(8401,-3061){\line( 0,-1){2400}}
\put(4801,-3061){\line( 0,-1){2400}}
\put(976,-6661){\line( 1, 0){225}}
\put(976,-7861){\line( 1, 0){5625}}
\put(6601,-3961){\line( 0,-1){3900}}
\put(1201,-6661){\circle{450}}
\put(12001,-361){\line( 0,-1){6300}}
\put(1276,-7711){\makebox(0,0)[lb]{\smash{\SetFigFont{7}{8.4}{rm}$-(\sqrt{13}+5)/2$}}}
\put(1201,-6661){\line( 1, 0){10800}}
\put(1201,-5761){\line( 1, 0){10800}}
\put(1201,-2761){\line( 1, 0){10800}}
\put(6451,-8011){\framebox(300,300){}}
\put(6451,-1711){\framebox(300,300){}}
\put(1051,-5911){\framebox(300,300){}}
\put(11851,-5911){\framebox(300,300){}}
\put(2851,-4111){\framebox(300,300){}}
\put(10051,-4111){\framebox(300,300){}}
\multiput(6826,-3961)(-7.24632,12.07721){17}{\makebox(11.1111,16.6667){\SetFigFont{7}{8.4}{rm}.}}
\put(6713,-3766){\line(-1, 0){224}}
\multiput(6489,-3766)(-7.24632,-12.07721){17}{\makebox(11.1111,16.6667){\SetFigFont{7}{8.4}{rm}.}}
\multiput(6376,-3961)(7.24632,-12.07721){17}{\makebox(11.1111,16.6667){\SetFigFont{7}{8.4}{rm}.}}
\put(6489,-4156){\line( 1, 0){224}}
\multiput(6713,-4156)(7.24632,12.07721){17}{\makebox(11.1111,16.6667){\SetFigFont{7}{8.4}{rm}.}}
\multiput(5026,-3061)(-7.24632,12.07721){17}{\makebox(11.1111,16.6667){\SetFigFont{7}{8.4}{rm}.}}
\put(4913,-2866){\line(-1, 0){224}}
\multiput(4689,-2866)(-7.24632,-12.07721){17}{\makebox(11.1111,16.6667){\SetFigFont{7}{8.4}{rm}.}}
\multiput(4576,-3061)(7.24632,-12.07721){17}{\makebox(11.1111,16.6667){\SetFigFont{7}{8.4}{rm}.}}
\put(4689,-3256){\line( 1, 0){224}}
\multiput(4913,-3256)(7.24632,12.07721){17}{\makebox(11.1111,16.6667){\SetFigFont{7}{8.4}{rm}.}}
\multiput(8626,-3061)(-7.24632,12.07721){17}{\makebox(11.1111,16.6667){\SetFigFont{7}{8.4}{rm}.}}
\put(8513,-2866){\line(-1, 0){224}}
\multiput(8289,-2866)(-7.24632,-12.07721){17}{\makebox(11.1111,16.6667){\SetFigFont{7}{8.4}{rm}.}}
\multiput(8176,-3061)(7.24632,-12.07721){17}{\makebox(11.1111,16.6667){\SetFigFont{7}{8.4}{rm}.}}
\put(8289,-3256){\line( 1, 0){224}}
\multiput(8513,-3256)(7.24632,12.07721){17}{\makebox(11.1111,16.6667){\SetFigFont{7}{8.4}{rm}.}}
\multiput(3226,-1261)(-7.24632,12.07721){17}{\makebox(11.1111,16.6667){\SetFigFont{7}{8.4}{rm}.}}
\put(3113,-1066){\line(-1, 0){224}}
\multiput(2889,-1066)(-7.24632,-12.07721){17}{\makebox(11.1111,16.6667){\SetFigFont{7}{8.4}{rm}.}}
\multiput(2776,-1261)(7.24632,-12.07721){17}{\makebox(11.1111,16.6667){\SetFigFont{7}{8.4}{rm}.}}
\put(2889,-1456){\line( 1, 0){224}}
\multiput(3113,-1456)(7.24632,12.07721){17}{\makebox(11.1111,16.6667){\SetFigFont{7}{8.4}{rm}.}}
\multiput(10426,-1261)(-7.24632,12.07721){17}{\makebox(11.1111,16.6667){\SetFigFont{7}{8.4}{rm}.}}
\put(10313,-1066){\line(-1, 0){224}}
\multiput(10089,-1066)(-7.24632,-12.07721){17}{\makebox(11.1111,16.6667){\SetFigFont{7}{8.4}{rm}.}}
\multiput(9976,-1261)(7.24632,-12.07721){17}{\makebox(11.1111,16.6667){\SetFigFont{7}{8.4}{rm}.}}
\put(10089,-1456){\line( 1, 0){224}}
\multiput(10313,-1456)(7.24632,12.07721){17}{\makebox(11.1111,16.6667){\SetFigFont{7}{8.4}{rm}.}}
\put(1201,-61){\line(-2,-3){300}}
\put(901,-511){\line( 1, 0){600}}
\put(1501,-511){\line(-2, 3){300}}
\put(1201,-61){\line( 0, 1){  0}}
\put(12001,-136){\line(-2,-3){300}}
\put(11701,-586){\line( 1, 0){600}}
\put(12301,-586){\line(-2, 3){300}}
\put(12001,-136){\line( 0, 1){  0}}
\put(10576,-8011){\framebox(300,300){}}
\multiput(7951,-7861)(-7.24632,12.07721){17}{\makebox(11.1111,16.6667){\SetFigFont{7}{8.4}{rm}.}}
\put(7838,-7666){\line(-1, 0){224}}
\multiput(7614,-7666)(-7.24632,-12.07721){17}{\makebox(11.1111,16.6667){\SetFigFont{7}{8.4}{rm}.}}
\multiput(7501,-7861)(7.24632,-12.07721){17}{\makebox(11.1111,16.6667){\SetFigFont{7}{8.4}{rm}.}}
\put(7614,-8056){\line( 1, 0){224}}
\multiput(7838,-8056)(7.24632,12.07721){17}{\makebox(11.1111,16.6667){\SetFigFont{7}{8.4}{rm}.}}
\put(7726,-6961){\line(-2,-3){300}}
\put(7426,-7411){\line( 1, 0){600}}
\put(8026,-7411){\line(-2, 3){300}}
\put(7726,-6961){\line( 0, 1){  0}}
\put(601,-361){\vector( 1, 0){11925}}
\put(976,-1786){\line( 1, 0){7425}}
\put(2851, 14){\makebox(0,0)[lb]{\smash{\SetFigFont{12}{14.4}{rm}$\pi/3$}}}
\put(4651, 14){\makebox(0,0)[lb]{\smash{\SetFigFont{12}{14.4}{rm}$2\pi/3$}}}
\put(6526, 14){\makebox(0,0)[lb]{\smash{\SetFigFont{12}{14.4}{rm}$\pi$}}}
\put(8251, 14){\makebox(0,0)[lb]{\smash{\SetFigFont{12}{14.4}{rm}$4\pi/3$}}}
\put(10051, 14){\makebox(0,0)[lb]{\smash{\SetFigFont{12}{14.4}{rm}$5\pi/3$}}}
\put(11851, 89){\makebox(0,0)[lb]{\smash{\SetFigFont{12}{14.4}{rm}$2\pi$}}}
\put(451, 14){\makebox(0,0)[lb]{\smash{\SetFigFont{17}{20.4}{rm}E/J}}}
\put(12226,-1036){\makebox(0,0)[lb]{\smash{\SetFigFont{17}{20.4}{rm}p}}}
\put(8101,-7261){\makebox(0,0)[lb]{\smash{\SetFigFont{9}{10.8}{rm}SEVENTHPLET}}}
\put(8101,-7945){\makebox(0,0)[lb]{\smash{\SetFigFont{9}{10.8}{rm}QUINTETS}}}
\put(11026,-7261){\makebox(0,0)[lb]{\smash{\SetFigFont{9}{10.8}{rm}TRIPLETS}}}
\put(11026,-7945){\makebox(0,0)[lb]{\smash{\SetFigFont{9}{10.8}{rm}SINGLETS}}}
\put(1201,-1486){\makebox(0,0)[lb]{\smash{\SetFigFont{7}{8.4}{rm}$-(5-\sqrt{13})/2$}}}
\put(1276,-1711){\makebox(0,0)[lb]{\smash{\SetFigFont{7}{8.4}{rm}$-(7-\sqrt{17})/2$}}}
\put(1201,-2011){\makebox(0,0)[lb]{\smash{\SetFigFont{7}{8.4}{rm}$-1$}}}
\put(1276,-1186){\makebox(0,0)[lb]{\smash{\SetFigFont{7}{8.4}{rm}$-1/2$}}}
\put(1276,-2536){\makebox(0,0)[lb]{\smash{\SetFigFont{7}{8.4}{rm}$-(5-\sqrt{5})/2$}}}
\put(1501,-2986){\makebox(0,0)[lb]{\smash{\SetFigFont{7}{8.4}{rm}$-3/2$}}}
\put(1276,-3886){\makebox(0,0)[lb]{\smash{\SetFigFont{7}{8.4}{rm}$-2$}}}
\put(1276,-4786){\makebox(0,0)[lb]{\smash{\SetFigFont{7}{8.4}{rm}$-5/2$}}}
\put(1276,-5386){\makebox(0,0)[lb]{\smash{\SetFigFont{7}{8.4}{rm}$-(\sqrt{17}+7)/4$}}}
\put(1501,-5686){\makebox(0,0)[lb]{\smash{\SetFigFont{7}{8.4}{rm}$-3$}}}
\put(1501,-6511){\makebox(0,0)[lb]{\smash{\SetFigFont{7}{8.4}{rm}$-(\sqrt{5}+5)/2$}}}
\end{picture}
\end{center}
\caption{Six sites chain spectrum}
\label{sixspectrum}
\end{figure}

Let us analize the spectrum of the 6 site antiferromagnetic chain. 
The momenta allowed for 
the states are now $0,\frac{\pi}{3},\frac{2\pi}{3},\pi,\frac{4\pi}{3},
\frac{5\pi}{3}\ {\rm mod}\  2\pi$. 
The ground state is
\begin{eqnarray}
|g.s.>&=&|0,0,-\frac{J}{2}(5+\sqrt{13}),\pi>=
\frac{1}{\sqrt{26-6\sqrt{13}}}
\{|\downarrow\uparrow\downarrow\uparrow\downarrow\uparrow>-
|\uparrow\downarrow\uparrow\downarrow\uparrow\downarrow>\nonumber\\
&+&\frac{1-\sqrt{13}}{6}(|\uparrow\uparrow\downarrow
\uparrow\downarrow\downarrow>-|\uparrow\downarrow\uparrow
\downarrow\downarrow\uparrow>+
|\downarrow\uparrow\downarrow\downarrow\uparrow\uparrow>-|
\uparrow\downarrow\downarrow\uparrow\uparrow\downarrow>
+|\downarrow\downarrow\uparrow\uparrow\downarrow\uparrow>- |
\downarrow\uparrow\uparrow\downarrow\uparrow\downarrow>\nonumber\\
&-&|\downarrow\downarrow\uparrow\downarrow\uparrow\uparrow>+|
\downarrow\uparrow\downarrow\uparrow\uparrow\downarrow>-
|\uparrow\downarrow\uparrow\uparrow\downarrow\downarrow>+|
\downarrow\uparrow\uparrow\downarrow\downarrow\uparrow>
-|\uparrow\uparrow\downarrow\downarrow\uparrow\downarrow>+|
\uparrow\downarrow\downarrow\uparrow\downarrow\uparrow>)\nonumber\\
&+&\frac{4-\sqrt{13}}{3}(|\uparrow\uparrow\uparrow\downarrow
\downarrow\downarrow>-|\uparrow\uparrow\downarrow\downarrow\downarrow\uparrow>+
|\uparrow\downarrow\downarrow\downarrow\uparrow\uparrow>
-|\downarrow\downarrow\downarrow\uparrow\uparrow\uparrow>+|
\downarrow\downarrow\uparrow\uparrow\uparrow\downarrow>-
|\downarrow\uparrow\uparrow\uparrow\downarrow\downarrow>)\}
\quad\quad\quad\quad .
\label{6gs}
\end{eqnarray}
This state is odd under $P$-parity.
The spectrum of the six sites chain is reported in 
fig.(\ref{sixspectrum}). There are 9 triplets in the spectrum. 
In \cite{pearson} 
it was already pointed out that the number of lowest 
lying triplets for a finite system with N sites is 
$N(N+2)/8$, so for $N=6$ there are 6 lowest lying triplet states. 
In order to identify these 6 states among the 9 that 
are exhibited by the spectrum of fig.(\ref{sixspectrum}), 
it is necessary to compute their $\lambda$'s and their $Q$'s.
In this way in fact, we can find out which are the triplets characterized by 
two holes and thus belonging to the triplet of type (\ref{c2}). In 
table (\ref{6tiqn}) we report the internal quantum numbers 
of the lowest lying triplets. The $Q^{(h)}$'s vary in the segment 
$[-\frac{3}{2},\frac{3}{2}]$.
The highest weight state of the triplet of zero 
momentum and energy $-(J/2)(5+\sqrt{5})$ reads
\begin{eqnarray}
|0,0,-\frac{J}{2}(5+\sqrt{5}),0>&=&\frac{1}{\sqrt{45-15\sqrt{5}}}
\{\frac{-3+\sqrt{5}}{2}(
|\downarrow\downarrow\uparrow\uparrow\uparrow\uparrow>+
|\downarrow\uparrow\uparrow\uparrow\uparrow\downarrow>+
|\uparrow\uparrow\uparrow\uparrow\downarrow\downarrow>+
|\uparrow\uparrow\uparrow\downarrow\downarrow\uparrow>\nonumber\\
&+&|\uparrow\uparrow\downarrow\downarrow\uparrow\uparrow>+
|\uparrow\downarrow\downarrow\uparrow\uparrow\uparrow>)\nonumber\\
&+&(|\downarrow\uparrow\downarrow\uparrow\uparrow\uparrow>+
|\uparrow\downarrow\uparrow\uparrow\uparrow\downarrow>+
|\downarrow\uparrow\uparrow\uparrow\downarrow\uparrow>+
|\uparrow\uparrow\uparrow\downarrow\uparrow\downarrow>+
|\uparrow\uparrow\downarrow\uparrow\downarrow\uparrow>+
|\uparrow\downarrow\uparrow\downarrow\uparrow\uparrow>)\nonumber\\
&+&(1-\sqrt{5})(|\downarrow\uparrow\uparrow\downarrow
\uparrow\uparrow>+|\uparrow\uparrow\downarrow\uparrow\uparrow\downarrow>+
|\uparrow\downarrow\uparrow\uparrow\downarrow\uparrow>)\}\quad\quad .
\label{tpo}
\end{eqnarray}
One can get the triplet of energy $-(J/2)(5-\sqrt{5})$ 
from (\ref{tpo}) by changing $\sqrt{5}\rightarrow -\sqrt{5}$.
As can be explicitly checked from (\ref{tpo}),
the two non-degenerate triplets of zero momentum are then $P$-parity even, 
namely they have opposite parity 
with respect to that of the  ground state, as it happens for the 
lowest lying triplet excitations of the 2-flavor 
Schwinger model. 
For what concerns the degenerate triplets of momenta 
$\pi/3$ and $5\pi/3$ (or $2\pi/3$ and $4\pi/3$) they do not have definite 
$P$-parity, but it is always possible to take a 
linear combination of them with parity opposite to the ground state.  
\begin{table}[htbp]
\begin{center}
\caption{Triplet internal quantum numbers }\label{6tiqn}
\vspace{.1in}
\begin{tabular}{|lcccc|}
\hline
TRIPLET& $\lambda_1$ & $\lambda_2 $ & $Q_{1}^{(h)}$ & 
$Q_{2}^{(h)}$ \rule{0in}{4ex}\\[2ex] \hline

$|1,1,-\frac{5+\sqrt{5}}{2}J,0>$ & $-\sqrt{\frac{5-2\sqrt{5}}{20}}$ 
& $\sqrt{\frac{5-2\sqrt{5}}{20}}$ & $-\frac{1}{2}$ & $\frac{1}{2}$ 
\rule{0in}{4ex}\\[2ex] \hline

$|1,1,-\frac{5-\sqrt{5}}{2}J,0>$ & $-\sqrt{\frac{5+2\sqrt{5}}{20}}$ 
& $\sqrt{\frac{5+2\sqrt{5}}{20}}$ & $-\frac{3}{2}$ & $\frac{3}{2}$     
\rule{0in}{4ex}\\[2ex] \hline

$|1,1,-\frac{5}{2}J,\frac{\pi}{3}>$ & $-\frac{\sqrt{3}+\sqrt{\pi}}{8}$ 
& $-\frac{\sqrt{3}-\sqrt{\pi}}{8}$ & $-\frac{3}{2}$ & $\frac{1}{2}$     
\rule{0in}{4ex}\\[2ex] \hline

$|1,1,-\frac{7+\sqrt{17}}{4}J,\frac{2\pi}{3}>$ & $\frac{-2\sqrt{3}-
\sqrt{-2+2\sqrt{17}}}{2+2\sqrt{17}}$ & 
$\frac{-2\sqrt{3}+\sqrt{-2+2\sqrt{17}}}{2+2\sqrt{17}}$ 
& $-\frac{3}{2}$ & $-\frac{1}{2}$     
\rule{0in}{4ex}\\[2ex] \hline

$|1,1,-\frac{7+\sqrt{17}}{4}J,\frac{4\pi}{3}>$ 
& $\frac{2\sqrt{3}-\sqrt{-2+2\sqrt{17}}}{2+2\sqrt{17}}$ & 
$\frac{2\sqrt{3}+\sqrt{-2+2\sqrt{17}}}{2+2\sqrt{17}}$ 
& $\frac{1}{2}$ & $\frac{3}{2}$     
\rule{0in}{4ex}\\[2ex] \hline

$|1,1,-\frac{5}{2}J,\frac{5\pi}{3}>$ & $\frac{\sqrt{3}-
\sqrt{\pi}}{8}$ & $\frac{\sqrt{3}+\sqrt{\pi}}{8}$ 
& $-\frac{1}{2}$ & $\frac{3}{2}$\rule{0in}{4ex}\\[2ex] \hline
\end{tabular}
\end{center}
\end{table} 

The remaining three triplets in fig.(\ref{sixspectrum}) 
have no real $\lambda$'s and are characterized by a 
string of length 2 and four holes 
for $Q=-\frac{3}{2},-\frac{1}{2},\frac{1}{2},\frac{3}{2}$, $i.e.$ do 
not belong to the type (\ref{c2}). More precisely, two 
triplets have a string approximately of length 2, due to the finite size 
of the system, while the triplet of momentum $\pi$ has a 
string exactly of length 2. In table (\ref{26tiqn}) we 
summarize the quantum numbers of 
these triplets. 

\begin{table}[htbp]
\begin{center}
\caption{Four holes triplet internal quantum numbers }\label{26tiqn}
\vspace{.1in}
\begin{tabular}{|lcc|}
\hline
TRIPLET&$\lambda_1$&$\lambda_2$\rule{0in}{4ex}\\[2ex] \hline

$|1,1,-\frac{7-\sqrt{17}}{4}J,\frac{2\pi}{3}>$ 
& $\frac{2\sqrt{3}-i\sqrt{2+2\sqrt{17}}}{-2+2\sqrt{17}}$ 
& $\frac{2\sqrt{3}+i\sqrt{2+2\sqrt{17}}}{-2+2\sqrt{17}}$ 
\rule{0in}{4ex}\\[2ex] \hline

$|1,1,-J,\pi>$ & $-\frac{i}{2}$ & $\frac{i}{2}$ 
\rule{0in}{4ex}\\[2ex] \hline

$|1,1,-\frac{7-\sqrt{17}}{4}J,\frac{4\pi}{3}>$ 
& $\frac{2\sqrt{3}+i\sqrt{2+2\sqrt{17}}}{2-2\sqrt{17}}$ 
& $\frac{2\sqrt{3}-i\sqrt{2+2\sqrt{17}}}{2-2\sqrt{17}}$ 
\rule{0in}{4ex}\\[2ex]\hline
\end{tabular}
\end{center}
\end{table} 

In fig.(\ref{sixspectrum}) it is shown that the spectrum 
exhibits five singlet states. The lowest lying state
at momentum $\pi$ is the ground state. Then there are 
three excited singlets characterized by the 
configuration with two holes (\ref{c1}), $i.e.$ they 
have one real $\lambda$ and a string of length almost 2.
In table (\ref{esin}) we summarize their quantum numbers. 
Among these singlets, those which are
not degenerate, have   
$P$-parity equal to that of the
ground state (odd) as it happens in the 2-flavor Schwinger 
model. The non-degenerate singlet in fact reads
\begin{eqnarray}
|0,0,-3J,0>&=&\frac{1}{\sqrt{12}}\{|\uparrow\uparrow\downarrow
\uparrow\downarrow\downarrow>+|\uparrow\downarrow\uparrow
\downarrow\downarrow\uparrow>+
|\downarrow\uparrow\downarrow\downarrow\uparrow\uparrow>+
|\uparrow\downarrow\downarrow\uparrow\uparrow\downarrow>+
|\downarrow\downarrow\uparrow\uparrow\downarrow\uparrow>+
|\downarrow\uparrow\uparrow\downarrow\uparrow\downarrow>\nonumber\\
&-&|\downarrow\downarrow\uparrow\downarrow\uparrow\uparrow>-
|\downarrow\uparrow\downarrow\uparrow\uparrow\downarrow>-
|\uparrow\downarrow\uparrow\uparrow\downarrow\downarrow>-
|\downarrow\uparrow\uparrow\downarrow\downarrow\uparrow>-
|\uparrow\uparrow\downarrow\downarrow\uparrow\downarrow>-
|\uparrow\downarrow\downarrow\uparrow\downarrow\uparrow>\}\quad \quad .\quad\quad
\label{spo}
\end{eqnarray}
The degenerate singlets are again not eigenstates of the $P$-parity, 
but it is always possible to take a linear combination of 
them with the a $P$-parity 
that coincides with that of the representative state (\ref{spo})
of the configuration.
\begin{table}[htbp]
\begin{center}
\caption{Singlet internal quantum numbers }\label{esin}
\vspace{.1in}
\begin{tabular}{|lcccc|}
\hline
SINGLET&$\lambda$ & $\lambda_S $ & $Q_1^{(h)}$ 
& $Q_2^{(h)}$\rule{0in}{4ex}\\[2ex] \hline

$|0,0,-3J,0>$ & 0 & 0 & $-1$ & 1 \rule{0in}{4ex}\\[2ex] \hline

$|0,0,-2J,\frac{\pi}{3}>$ & $-\frac{\sqrt{3}+2\sqrt{6}}{14}$ 
& $\frac{-2+3\sqrt{2}}{\sqrt{3}(4+\sqrt{2})}$ & $0$ 
& 1 \rule{0in}{4ex}\\[2ex] \hline

$|0,0,-2J,\frac{5\pi}{3}>$ & $\frac{\sqrt{3}+2\sqrt{6}}{14} $ 
&  $\frac{2-3\sqrt{2}}{\sqrt{3}(4+\sqrt{2})} $& $-1$ 
& 0 \rule{0in}{4ex}\\[2ex]\hline
\end{tabular}
\end{center}
\end{table}

The remaining singlet $|0,0,-\frac{5-\sqrt{13}}{2} J,\pi>$ 
it is not of the type (\ref{c1}). It is characterized by 
a string approximately of length 3 with $\lambda_{1,1}=
i\sqrt{\frac{5+2\sqrt{13}}{12}}$, $\lambda_{2,1}=0$ 
and $\lambda_{3,1}=-i\sqrt{\frac{5+2\sqrt{13}}{12}}$.

Even in finite systems very small like the 4 and 6 sites chains, 
the ``string hypothesis" is a very good approximation and 
it allows us to classify and distinguish among states with the same spin.   

The ground state of the antiferromagnetic Heisenberg 
chain with N sites is a linear combination of all the 
$\left( \begin{array}{c} N \\ 
\frac{N}{2} 
\end{array} \right)$ states with $\frac{N}{2}$ spins up and $\frac{N}{2}$ 
spins down. 
These states group themselves into sets with the same coefficient 
in the linear combination  according to the fact that 
the ground state is translationally invariant (with momentum 0 ($\pi$) 
for $\frac{N}{2}$ even (odd)), it 
is an eigenstate of $P$-parity and it is 
invariant under the exchange of up with down spins. 
The states belonging to the same set 
have the same number of domain walls, which ranges from 
$N$, for the two N\'eel states, to 2 for the states with $\frac{N}{2}$ 
adjacent spins up and $\frac{N}{2}$ adjacent spins down.  

The ground state of the 8 sites chain is 

\begin{equation}
|g.s.>= \frac{1}{ \sqrt{\cal{N}} }
(|\psi_{8} >+\alpha |\psi_6 ^{(1)} >+\beta |\psi_6^{(2)} >
+\gamma |\psi_4 ^{(1)}>+\delta |\psi_4 ^{(2)}>+\epsilon
|\psi_4 ^{(3)} >+\zeta |\psi_2>)
\label{8gs}
\end{equation} 
where 
\begin{eqnarray}
& &|\psi_{8}> = |\uparrow\downarrow\uparrow\downarrow\uparrow
\downarrow\uparrow\downarrow>+
|\downarrow\uparrow\downarrow\uparrow\downarrow\uparrow
\downarrow\uparrow>\\
& &|\psi_6 ^{(1)}> = |\uparrow\uparrow\downarrow\uparrow
\downarrow\uparrow\downarrow\downarrow>+
|\downarrow\downarrow\uparrow\downarrow\uparrow\downarrow
\uparrow\uparrow>+translated\quad states \\
& &|\psi_6^{(2)}>= |\uparrow\uparrow\downarrow\uparrow
\downarrow\downarrow\uparrow\downarrow>+
|\downarrow\downarrow\uparrow\downarrow\uparrow\uparrow
\downarrow\uparrow>+translated\quad states\\
& &|\psi_4 ^{(1)}>= |\uparrow\uparrow\downarrow\downarrow
\uparrow\uparrow\downarrow\downarrow>+translated\quad states\\
& &|\psi_4 ^{(2)}> = |\uparrow\uparrow\uparrow\downarrow
\downarrow\downarrow\uparrow\downarrow>+
|\downarrow\downarrow\downarrow \uparrow\uparrow\uparrow
\downarrow\uparrow>+translated\quad states \\
& &|\psi_4 ^{(3)}> = |\uparrow\uparrow\downarrow\downarrow
\downarrow\uparrow\uparrow\downarrow>+
|\downarrow\downarrow\uparrow\uparrow\uparrow\downarrow
\downarrow\uparrow>+translated\quad states\\
& &|\psi_2> = |\uparrow\uparrow\uparrow\uparrow\downarrow
\downarrow\downarrow\downarrow>+translated\quad states\ .
\label{8states}
\end{eqnarray}
By direct diagonalization one gets
\begin{eqnarray}
\alpha&=&-0.412773\\
\beta&=&0.344301\\
\gamma&=&0.226109\\
\delta&=&-0.087227\\
\epsilon&=&0.136945\\
\zeta&=&0.018754\\
\cal{N}&=&2+16\alpha^2+8\beta^2+4\gamma^2+16\delta^2
+16\epsilon^2+8\zeta^2=6.30356\quad .
\end{eqnarray}

The energy of the ground state is 
\begin{equation}
E_{g.s.}=-5.65109J\quad .
\label{egs8}
\end{equation}
Eq.(\ref{egs8}) differs only by $1.8\%$ from the thermodynamic 
limit expression $E_{g.s.}=-8\ln 2=-5.54518$. Moreover 
also the correlation function of distance 2 Eq.(\ref{corrd2}) 
computed for the 8 sites chain is $G(2)=0.1957N$, value which is $7\%$ 
higher than the exact answer Eq.(\ref{corrd2n}).

In the analysis of finite size systems we were able to find 
the coefficient $\beta$ of the first set of states containing
$N-2$ domain walls in 
the ground state. These states 
are obtained interchanging two adjacent spins in the N\'eel states. 
The $\beta$ is for a generic chain of $N$-sites
\begin{equation}
\beta=\frac{N+2E_{g.s.}}{N}=1-2\ln 2\quad .
\end{equation}    

\noindent
{\Large{\bf Appendix B: The correlator $G(2)$ in terms of spin 
configuration probabilities}}

In this appendix we shall establish a relation between the Heisenberg model
correlator $G(2)$ and the probabilities of finding,
in the antiferromagnetic vacuum, certain groups of 
spin in a given position.

The isotropy of the Heisenberg model implies that
\begin{equation}
\sum_{x=1}^{N}<g.s.|\vec{S}_{x}\cdot \vec{S}_{x+2}|g.s.>=3
\sum_{x=1}^{N}<g.s.|S^{3}_{x}\cdot S^3_{x+2}|g.s.>\quad .
\label{e1}
\end{equation}
Let us introduce the probability $P_3$ for finding three adjacent spins
in a given position in the Heisenberg antiferromagnetic vacuum. 
Taking advantage of the isotropy of the Heisenberg model 
ground state and of its
translational invariance, it is easy to see that the correlator (\ref{e1}) 
can be written in terms of the $P_3$'s as
\begin{equation}
\sum_{x=1}^{N}<g.s.|S^{3}_{x}\cdot S^3_{x+2}|g.s>=N~
\frac{1}{4}~2~(~P_3(\uparrow\uparrow\uparrow)+P_3(\uparrow\downarrow\uparrow)-
P_3(\uparrow\uparrow\downarrow)-P_3(\downarrow\uparrow\uparrow)~)\quad .
\label{e2}
\end{equation}
The factor 2 appears in (\ref{e2}) due again to the isotropy of the 
Heisenberg model: the probability of  
a configuration and of the configuration rotated by $\pi$ around the 
chain axis, are the same.  

In \cite{korepin} the so called ``emptiness formation probability'' $P(x)$
was introduced.
\begin{equation}
P(x)=<g.s.|\prod_{j=1}^{x}P_{j}|g.s.>\quad ,
\label{e3}
\end{equation}
where 
\begin{equation}
P_{j}=\frac{1}{2}(\sigma_{j}^{3}+1)
\label{e4}
\end{equation}
and $\sigma_{j}^{3}$ is the Pauli matrix.
$P(x)$ determines the probability of finding $x$ adjacent spins up in the 
antiferromagnetic vacuum.
One gets
\begin{eqnarray}
P(\uparrow\uparrow\uparrow)&=&P(3)\\
P(\uparrow\downarrow\uparrow)&=&P(1)-2P(2)+P(3)\\
P(\downarrow\uparrow\uparrow)&=&P(\uparrow\uparrow\downarrow)=P(2)-P(3)
\end{eqnarray}
so that Eq.(\ref{corrd2}) reads
\begin{equation}
\frac{G(2)}{3}=2P(3)-2P(2)+\frac{1}{2}P(1)\quad .
\label{corrd22}
\end{equation}
Using the exact  value the correlator $G(2)$ computed in~\cite{taka}
from (\ref{corrd22}) and from the known values of $P(2)$
and $P(1)$ given in \cite{korepin}
\begin{eqnarray}
P(1)&=&\frac{1}{2}\label{p1}\\
P(2)&=&\frac{1}{3}(1-\ln 2)\label{p2}\\
\label{p1p2}
\end{eqnarray}
we get\footnote{We thank V. Korepin for pointing out a misprint in this formula
in the previous version of this paper.}
\begin{eqnarray}
P(3)&=&\frac{1}{4}-\ln 2+\frac{3}{8} \zeta(3)\label{p3}\quad .
\label{efp3}
\end{eqnarray}
For the general emptiness formation probability $P(x)$ of the antiferromagnetic
Heisenberg chain,
an integral representation was given in~\cite{korepin},
but, to our knowledge, the exact value of $P(3)$ (\ref{efp3}) was not
known.

\end{document}